%% file: ms.tex
\documentclass{emulateapj}

\usepackage{apjfonts}
\usepackage{amssymb}
\usepackage{epsfig}
\usepackage{amsmath}
\usepackage{graphicx} 
\usepackage{relsize}
\usepackage[toc,page]{appendix}
\usepackage[T1]{fontenc}
\usepackage{chngcntr}
\usepackage{array} 
\usepackage{verbatim}

\usepackage[backref,breaklinks,colorlinks,citecolor=blue]{hyperref}
\usepackage[all]{hypcap}

\usepackage[caption = false]{subfig}

\newcommand{\silicate}{$S_{9.7\,\mu\mathrm{m}}\,$}

\begin{document}

\title{Search for Molecular Outflows in Local Volume AGN with \emph{Herschel}-PACS\footnotemark[1]\\}
\footnotetext[1]{\emph{Herschel} is an ESA space observatory with science instruments provided by European-led Principal Investigator consortia and with important participation from NASA.}

\author{M. Stone, S. Veilleux\footnotemark[2], M. Mel\'{e}ndez\footnotemark[3,]\footnotemark[4]} 
\affil{Department of Astronomy, University of  Maryland, College Park, MD 20742-2421; \\ mjstone@astro.umd.edu, veilleux@astro.umd.edu, marcio@astro.umd.edu }
\footnotetext[2]{Joint Space-Science Institute, University of Maryland, College Park, MD 20742, USA}
\footnotetext[3]{NASA Goddard Space Flight Center, Greenbelt, MD 20771}
\footnotetext[4]{Wyle Science, Technology and Engineering Group, 1290 Hercules Avenue, Houston, TX 77058}

\author{E. Sturm, J. Graci\'{a}-Carpio}
\affil{Max-Planck-Institute for Extraterrestrial Physics (MPE), Giessenbachstrasse 1, D-85748 Garching, Germany\\}  

\author{E. Gonz\'{a}lez-Alfonso}
\affil{Universidad de Alcal\'a, Departamento de F\'{\i}sica y Matem\'{a}ticas, Campus Universitario, E-28871 Alcal\'a de Henares, Madrid, Spain}
 
\begin{abstract}
We present the results from a systematic search for galactic-scale, molecular (OH 119 $\mu$m) outflows in a sample of 52 Local Volume ($d < 50$ Mpc) Burst Alert Telescope detected active galactic nuclei (BAT AGN) with \emph{Herschel}-PACS. We  combine the results from our analysis of the BAT AGN with the published \emph{Herschel}/PACS data of 43 nearby ($z<0.3$) galaxy mergers, mostly ultraluminous infrared galaxies (ULIRGs) and QSOs. The objects in our sample of BAT AGN have, on average, $\sim 10-100$ times lower AGN luminosities, star formation rates (SFRs), and stellar masses than those of the ULIRG and QSO sample. OH 119 $\mu$m is detected in 42 of our BAT AGN targets. Evidence for molecular outflows (i.e. OH absorption profiles with median velocities more blueshifted than $-$50 km s$^{-1}$ and/or blueshifted wings with 84-percentile velocities less than $-$300 km s$^{-1}$) is seen in only four BAT AGN (NGC~7479 is the most convincing case). Evidence for molecular inflows (i.e. OH absorption profiles with median velocities more redshifted than 50 km s$^{-1}$) is seen in seven objects, although an inverted P-Cygni profile is detected unambiguously in only one object (Circinus). Our data show that both the starburst and AGN contribute to driving OH outflows, but the fastest OH winds require AGN with quasar-like luminosities. We also confirm that the total absorption strength of OH 119 $\mu$m is a good proxy for dust optical depth as it correlates strongly with the 9.7 $\mu$m silicate absorption feature, a measure of obscuration originating in both the nuclear torus and host galaxy disk. \\
\end{abstract}

\section{Introduction}

Massive, galactic-scale outflows driven by star formation and/or active galactic nuclei (AGN) may be the dominant form of feedback in galaxies \citep{Veilleux2005, Fabian2012}. These outflows (or winds) likely affect the evolution of galaxies by regulating star formation and black hole (BH) accretion activity. These winds may shut off the gas feeding process and stop the growth of both the BH and the spheroidal component \citep{diMatteo2005}, thereby explaining the tight ``bulge-BH mass relation" \citep[e.g.][]{Fabian2012, Marconi2003, Kormendy1995, Kormendy2013}. They may also quench star formation altogether and help explain the presence of ``red-and-dead", gas-poor ellipticals, and the bimodal color distribution observed in large galaxy surveys \citep[e.g.][]{Strateva2001, Baldry2004}. 
Winds may also be the primary mechanism by which metals are transferred from galaxies to their surrounding halos, and to a lesser extent, to the intergalactic medium.
\input{tab1}
\begin{figure*}[t!]
\label{fig:distributions}
\centering
\includegraphics[scale=0.9]{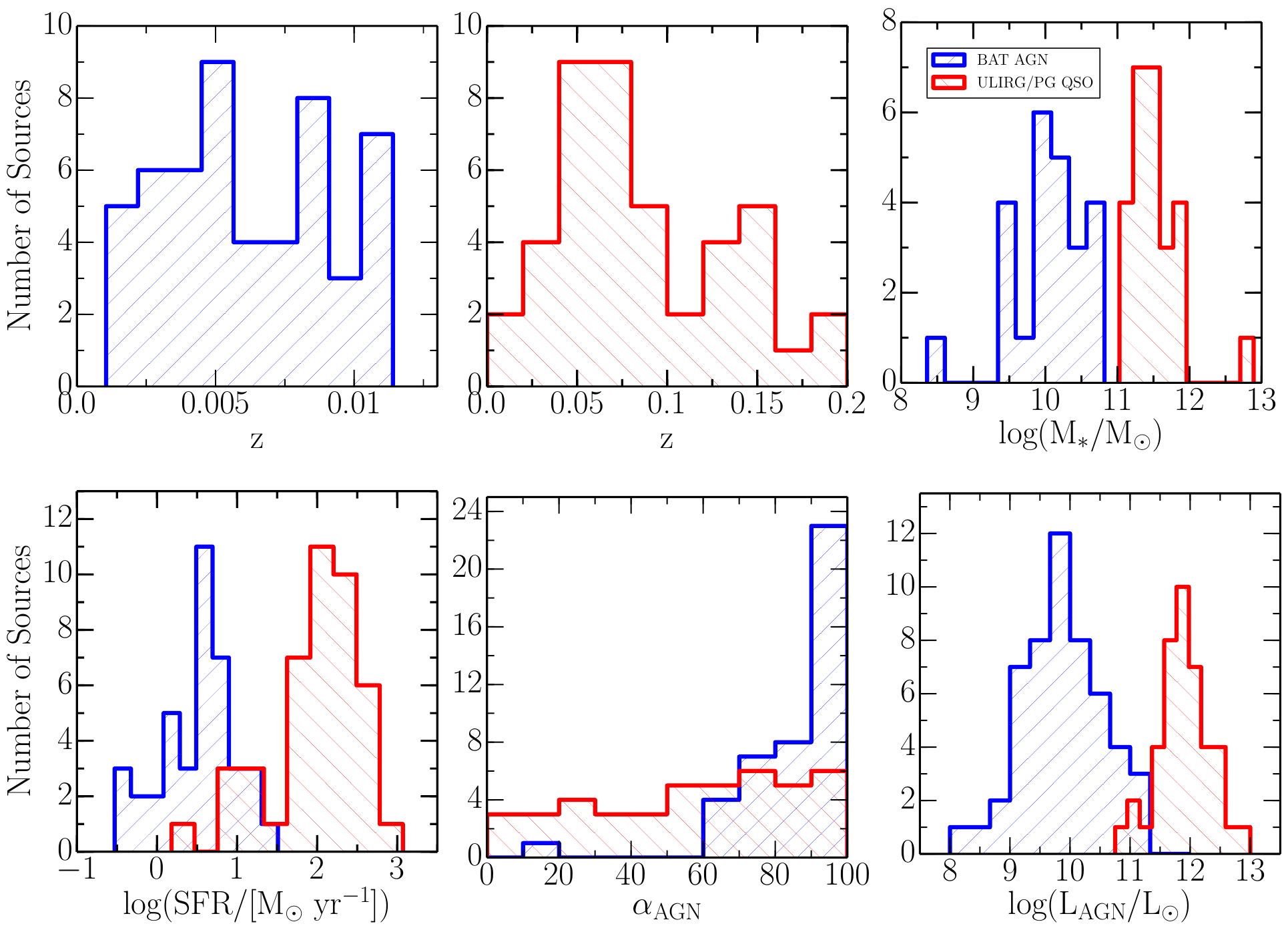}
\caption{Histograms showing the distributions of the BAT AGN and ULIRG/PG QSO properties: redshifts, stellar masses, star formation rates (SFR), AGN fractions, and AGN luminosities.}
\end{figure*}
Until recently, searches for galactic-scale outflows have focused on the brightest sources in bands often affected by obscuration and/or contamination from the host galaxy light \citep[e.g.][]{Lehnert1996}. These searches have generally been directed at either the ionized phase \citep[e.g.][]{Crenshaw1999, Dunn2008, Rubin2010} or the neutral phase \citep[e.g.][]{Heckman2000, Schwartz2004, RupkeFirst2005, RupkeSecond2005, RupkeThird2005, Martin2005, Krug2010, Rupke2011, Rupke2013} of the ISM. If winds are to inhibit star formation in the host galaxy, then the mass outflow must affect the phase of the ISM from which stars form (i.e. the cold molecular gas). Our knowledge of molecular outflows is quickly improving. Recent studies of galactic-scale winds have effectively demonstrated that far infrared (FIR) spectroscopy of the hydroxyl molecule (OH) with \emph{Herschel}-PACS is well suited to identify molecular outflows in the nearby universe \citep[][hereafter, S11, V13, and S13 respectively]{Sturm2011, Veilleux2013, Spoon2013}. S11 reported preliminary evidence of a correlation between AGN luminosity ($L_{AGN}$) and OH terminal velocity in six ultra-luminous infrared galaxies (ULIRGs; i.e. ULIRGs with higher terminal velocities hosted AGN with higher luminosities). V13 and S13 later confirmed this correlation via the analyses of larger samples (43 and 24 ULIRGs, respectively). In particular, V13 reported a nonlinear relationship between log($L_{AGN}/L_\odot$) and outflow velocity, but noted that better statistics were required at lower AGN luminosities in order to confirm this nonlinearity. These \emph{Herschel}-based studies, supplemented with millimeter-wave, interferometric studies, have also shown that the molecular gas often dominates the mass and energy budget of these outflows \citep[e.g.][]{Fischer2010, Feruglio2010, Alatalo2011, Sturm2011, Morganti2013, Veilleux2013, Gonzalez2014, Cicone2014}. 

There is thus a clear need to extend this type of study to lower AGN luminosities and star formation rates (SFRs). For this, we examine \emph{Herschel} observations of a complete sample of local \emph{Swift}-BAT selected AGN. Since stellar processes contribute negligibly to the 14-195 keV emission, the BAT AGN survey is not sensitive to star formation activity within the host galaxy. Additionally, this survey is unbiased to column densities of $N_H \lesssim 10^{24}$ cm$^{-2}$. The characteristics and host galaxy properties of these ultra-hard X-ray detected BAT AGN have been studied extensively across most wavelengths \citep[e.g.][]{Winter2010, Vasudevan2009, Vasudevan2010, Koss2011, Koss2013, Matsuta2012, Mushotzky2014, Melendez2014, Shimizu2015, Shimizu2016}. By combining the results of our analysis on 52 BAT AGN with those of V13 on 43 ULIRGs, we extend the range of AGN luminosities, SFRs, and stellar masses sampled in this study by 1-2 orders of magnitude, from which we can draw stronger statistical conclusions on the driving mechanisms of these outflows. We also examine mid-infrared (MIR; 5-37 $\mu$m) spectra from the Infrared Spectrograph (IRS) on board the \emph{Spitzer Space Telescope} of the combined BAT AGN + ULIRG + PG QSO sample in an attempt to constrain the distributions of the dust, as measured by the strength of the 9.7 $\mu$m silicate feature, and OH gas within these systems. We note that although robust, statistical conclusions can only be drawn from a well defined sample, valuable insight into molecular winds may still be obtained by combining these two samples which are distinct in their selection methods and in their properties.

\begin{figure*}[t!]
\label{fig:ohSpectralFits}
\centering
\includegraphics[scale=0.94]{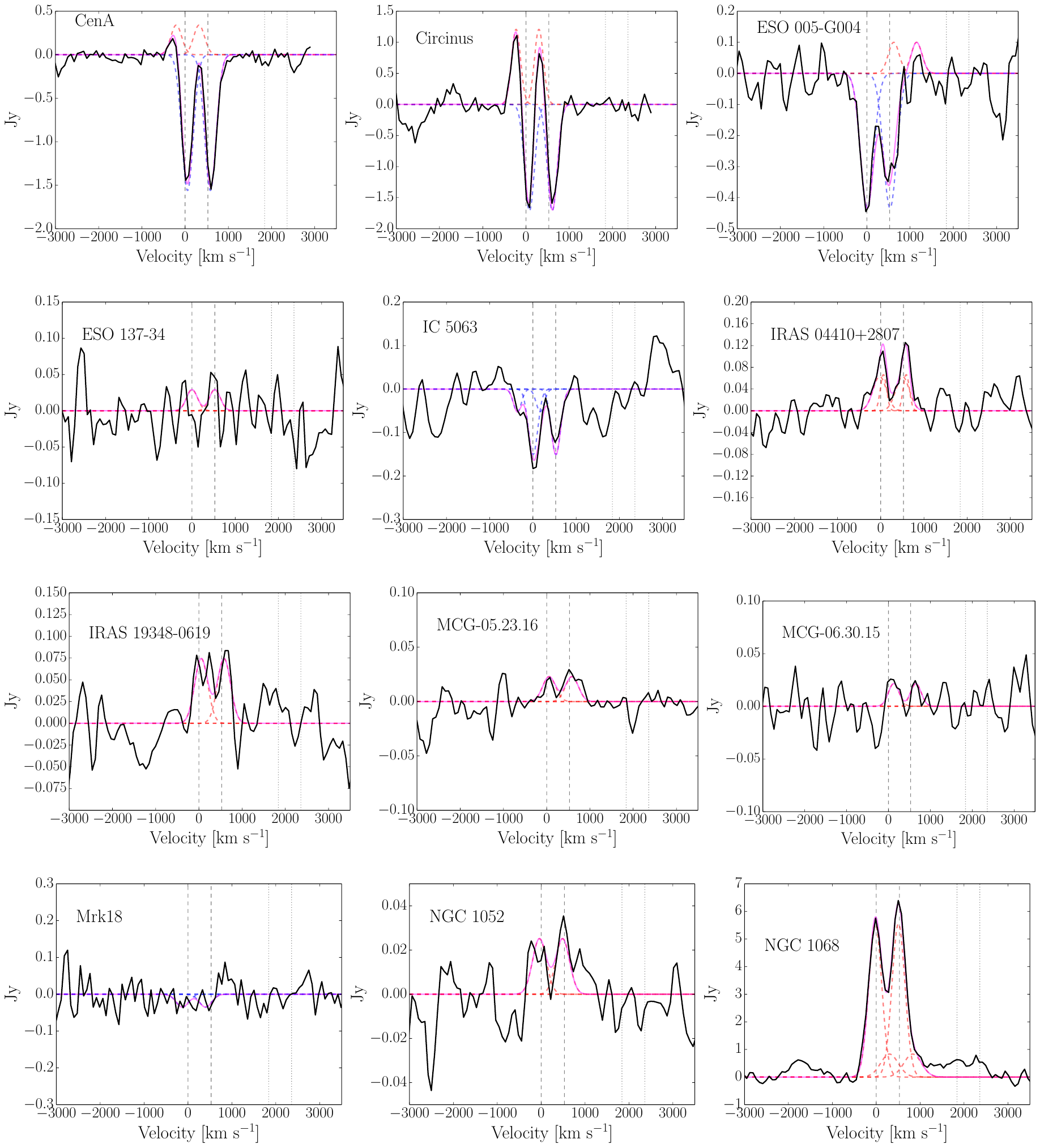}
\caption{Fits to the central spaxel, continuum subtracted OH 119 $\mu$m profiles of the 52 objects in our sample; see \autoref{sec:ohAnalysis}. In each figure, the solid black line represents the data and the magenta line is the best fit to the data. The blue and red dashed lines represent the Gaussian components of the fits. The origin of the velocity scale corresponds to OH 119.233 $\mu$m at the systemic velocity. The two vertical dashed and dotted lines mark the positions of $^{16}$OH and $^{18}$OH, respectively.}
\end{figure*}
\setcounter{figure}{1}
\begin{figure*}[t!]
\label{fig:ohSpectralFits}
\centering
\includegraphics[scale=0.94]{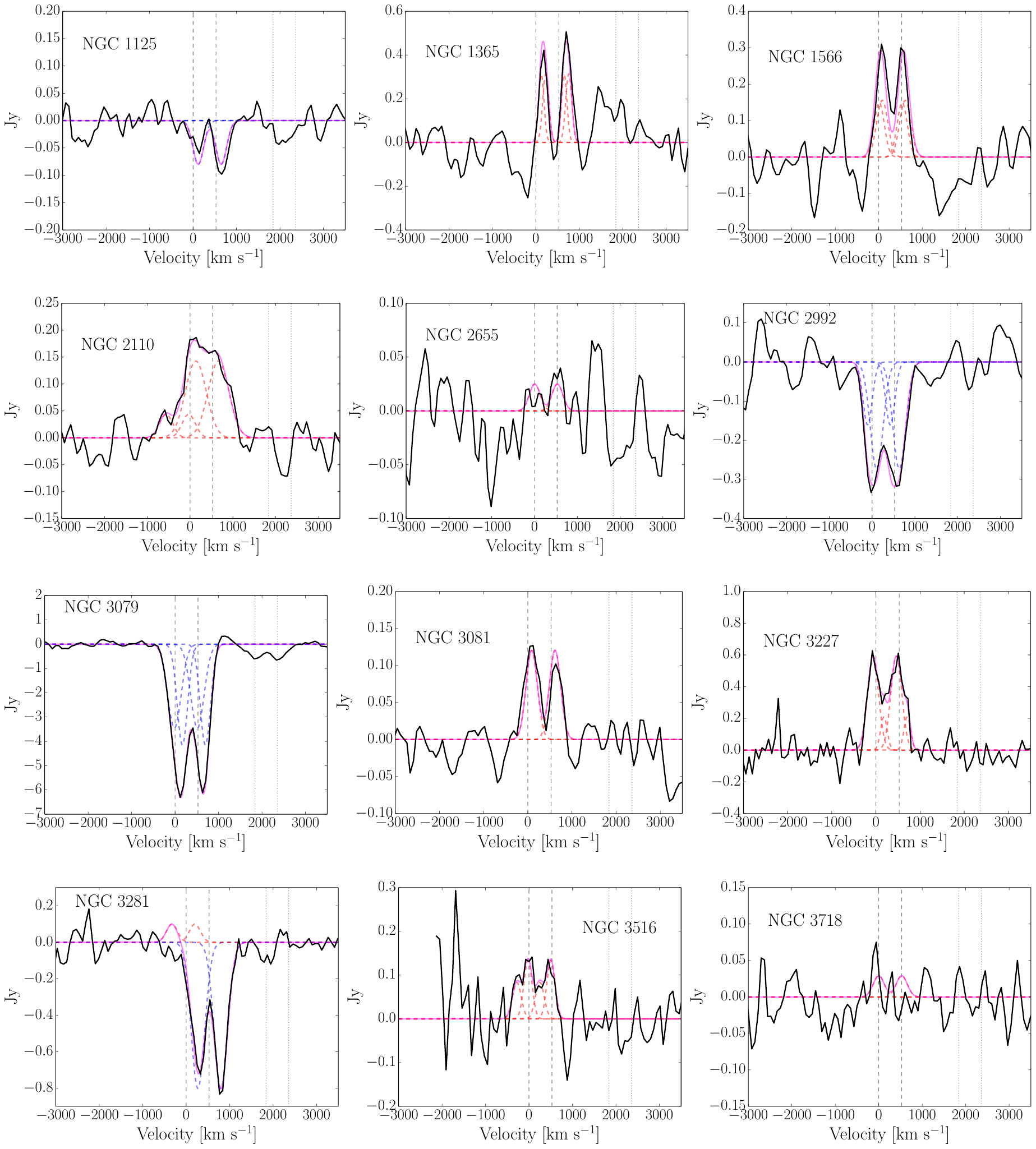}
\caption{(Continued)}
\end{figure*}
\setcounter{figure}{1}
\begin{figure*}[t!]
\label{fig:ohSpectralFits}
\centering
\includegraphics[scale=0.94]{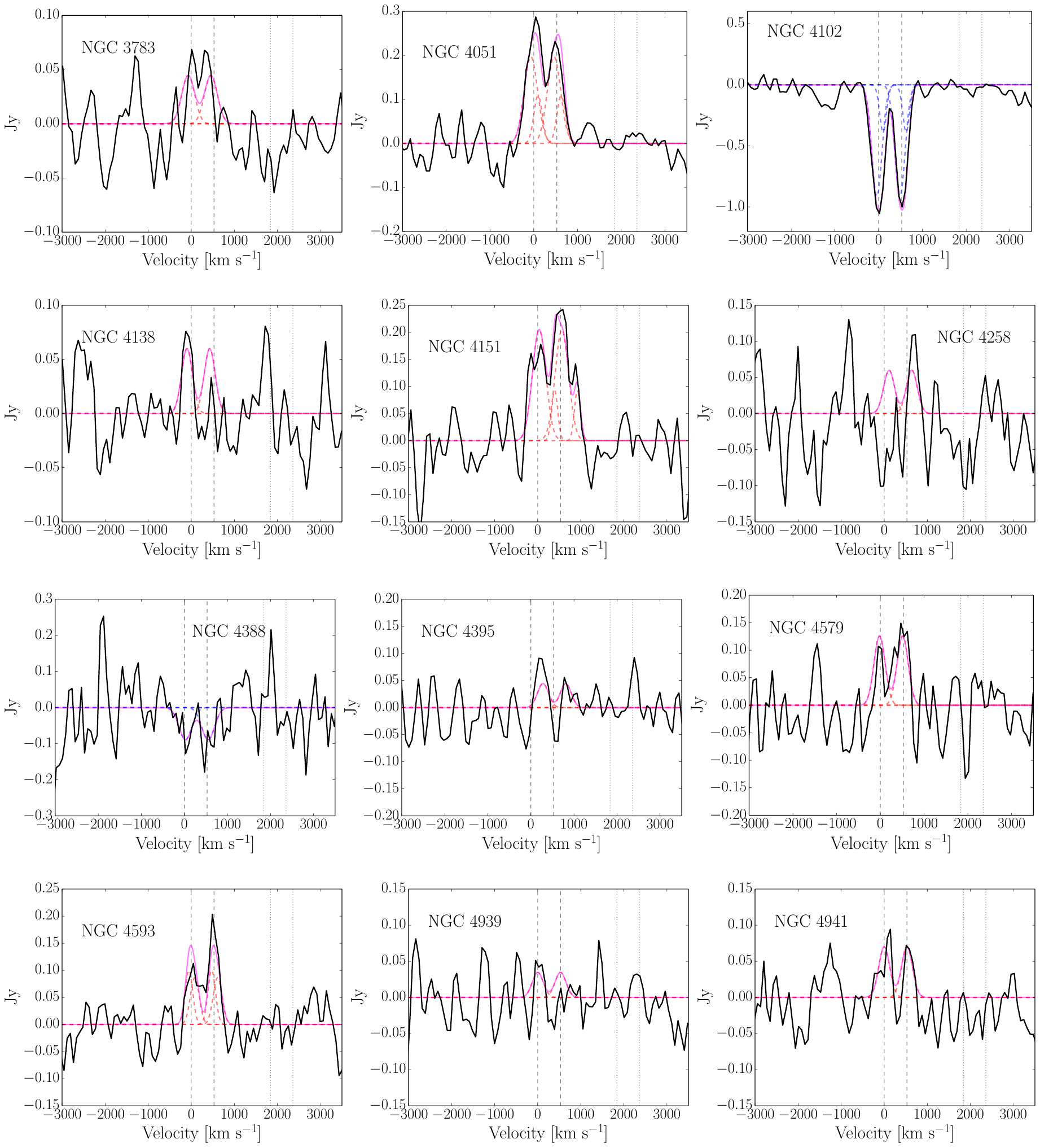}
\caption{(Continued)}
\end{figure*}
\setcounter{figure}{1}
\begin{figure*}[t!]
\label{fig:ohSpectralFits}
\centering
\includegraphics[scale=0.94]{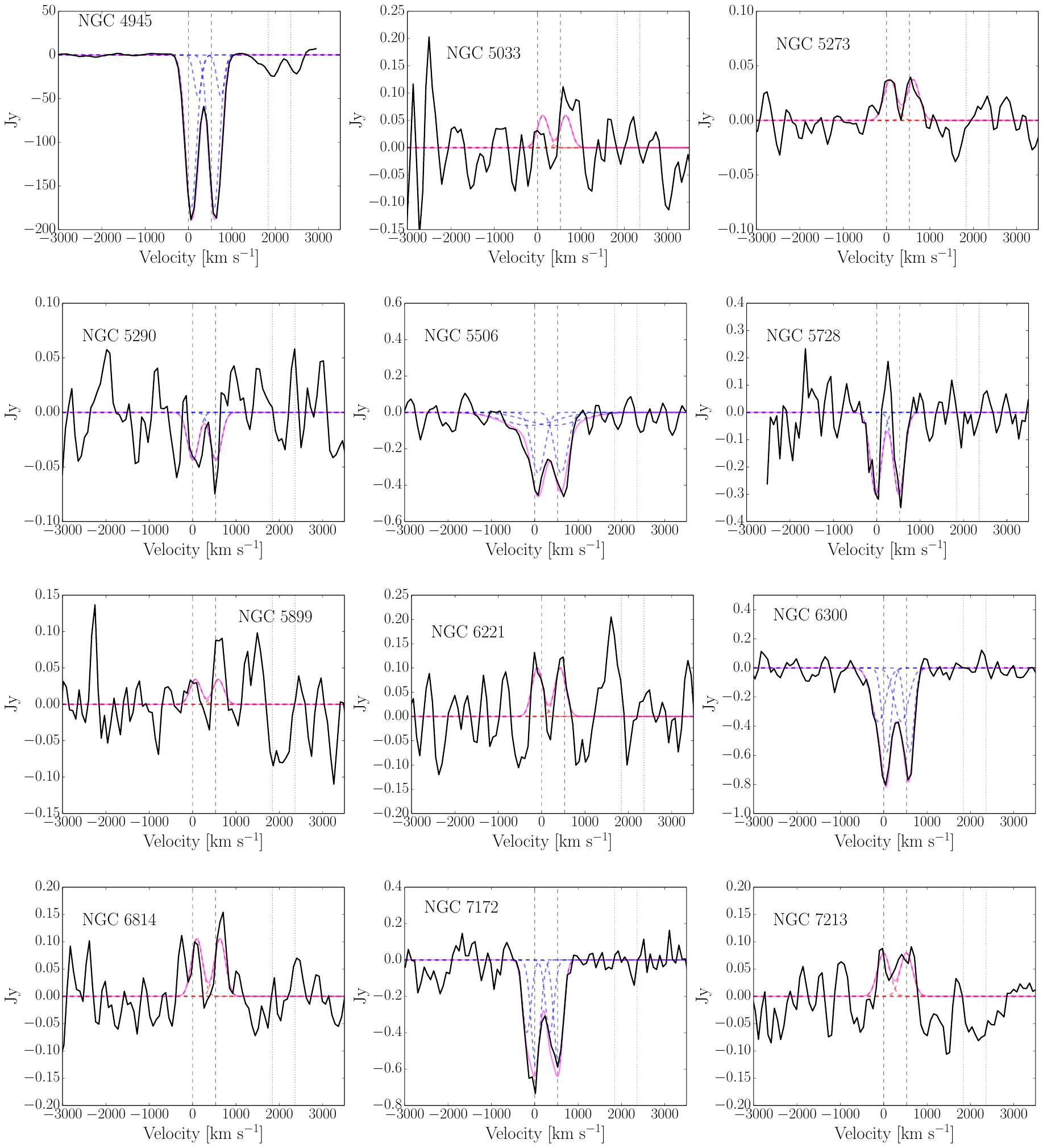}
\caption{(Continued)}
\end{figure*}
\setcounter{figure}{1}
\begin{figure*}[t!]
\label{fig:ohSpectralFits}
\centering
\includegraphics[scale=0.94]{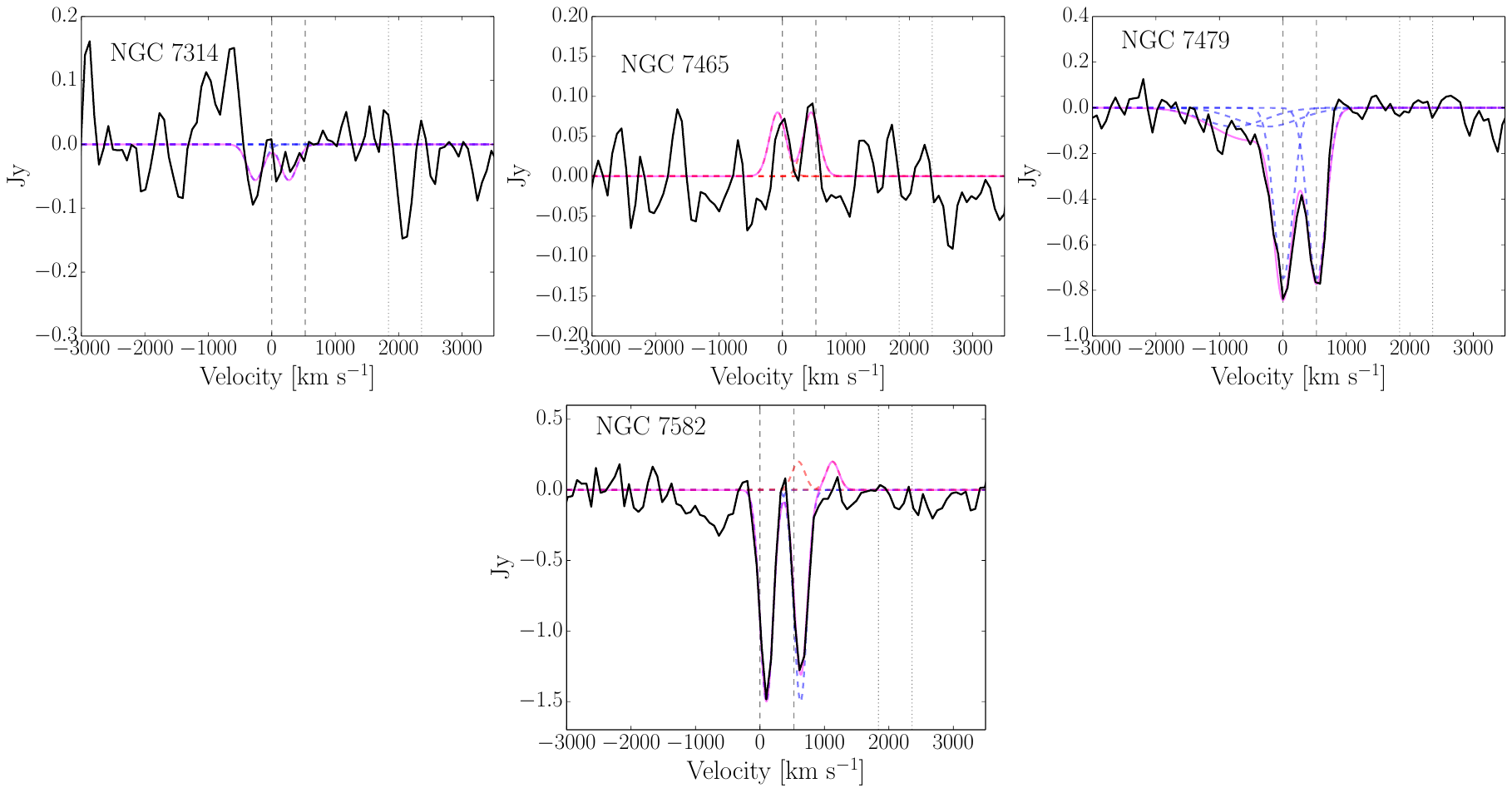}
\caption{(Continued)}
\end{figure*}

The samples used in this analysis are described in Section 2. The observations and data reduction techniques are outlined in Section 3. The results of the analysis are presented in Section 4, while the implications of these results are reported in Section 5. The conclusions are summarized in Section 6.

\section{Sample}
\subsection{BAT AGN Sample Selection}

The BAT AGN in our sample were selected using three criteria: (1) all targets are from the very hard X-ray selected (14-195 keV) 58-month \emph{Swift}-BAT Survey \citep{Baumgartner2011} of local AGN. Since the 14-195 keV flux is solely produced by the AGN and unaffected by the host galaxy light or obscuration along the line-of-sight ($N_H \lesssim 10^{24}$ cm$^{-2}$, this criterion removes any ambiguity as to the power of the AGN. Thus, the \emph{Swift}-BAT survey is arguably superior to soft X-ray, UV, optical, IR, or radio surveys for understanding the role of AGN-driven winds in galaxies. (2) All targets have a total integrated flux at 120 $\mu$m of $S_{120}^{\mathrm{tot}} \gtrsim 1$ Jy so that high S/N in the continuum can be reached in a reasonable amount of time with PACS. (3) Finally, all targets are located within 50 Mpc. The low redshifts in this sample provide the best possible scale ($\sim$ 0.02-0.2 kpc arcsec$^{-1}$) for spatially separating star-formation emission from the nuclear component. This distance is also large enough to properly sample the AGN luminosity function up to quasar-like values (log $L_{\mathrm{BAT}} \sim 43.5$; log $L_{\mathrm{BOL}}\sim45$), without favoring IR-bright systems due to criterion \#2. 

We find that 52 targets meet these three requirements. Of these targets, 42 objects are from the cycle 2 open-time program OT2\_sveilleu\_6 (PI: S. Veilleux), 3 objects are from the guaranteed time program GT1\_lspinogl\_4 (PI: L. Spinoglio), 2 objects are from the guaranteed time program GT1\_lspinogl\_6 (PI: L. Spinoglio), 3 objects are from the cycle 1 open-time program OT1\_shaileyd\_1 (PI: S. Hailey-Dunsheath), 1 object is from the Director's Discretionary Time DDT\_esturm\_4 (PI: E. Sturm), and 1 object is from the guaranteed time key program Survey with Herschel of the ISM in Nearby Infrared Galaxies (SHINING; PI: E. Sturm) (see \autoref{tab:batAgnObservations}). 

\input{tab2}

\subsection{ULIRG and PG QSO Sample}
We include in our analysis the \emph{Herschel}/PACS spectra of the ULIRGs and PG QSOs from V13. Of the 43 objects (38 ULIRGs + 5 QSOs) from that sample, 23 targets are from the key program SHINING (PI: E. Sturm), 15 targets are from the cycle 1 open-time program OT1\_sveilleu\_1 (PI: S. Veilleux), and 5 are from the cycle 2 open-time program OT2\_sveilleu\_4 (PI: S. Veilleux).

This sample spans a broad range of merger stages. 20 objects are cool ($f_{25}/f_{60} \leq 0.2$ ) pre-merger ULIRGs, 18 objects are warm ($f_{25}/f_{60} > 0.2$) quasar-dominated, late-stage, fully coalesced ULIRGs, and 5 objects are ``classic" IR-faint QSOs. These QSOs are in a late merger phase in which the quasar has finally shed its natal ``cocoon" of dust and gas and feedback effects may be receding \citep[][V13]{Veilleux2009}

\subsection{Properties of the Sample Galaxies}\label{sec:galaxyProperties}

The properties of our BAT AGN sample are listed in \autoref{tab:batAgnProperties}. The notes to \autoref{tab:batAgnProperties} briefly explain the meaning of each of these quantities. Some quantities, however, require further clarification. We apply the bolometric correction from \citet{Winter2012} to our BAT AGN luminosities, which are derived from the fluxes from the \emph{Swift} BAT 70-month survey. Since the \emph{Swift}-BAT bandpass is at high enough energies (14-195 keV) to be unaffected by all but the highest levels of obscuration, the luminosities in this bandpass should be the direct unobscured signature from the AGN. Thus, it is assumed to be a good proxy for the bolometric luminosity of the AGN. The correction is a scale factor derived from the correlation between the bolometric luminosity, which is determined from simultaneous broadband fitting of the spectral energy distribution of 33 sources in the optical, UV, and X-ray \citep{Vasudevan2007, Vasudevan2009}, and the \emph{Swift} BAT band 14-195 keV luminosities of those sources. The ordinary least-squares line through these data of \citet{Winter2012} yields the following correction:
\begin{equation} \label{eq:bolCorrection}
L_{\mathrm{AGN}}=10.5\times L_{\mathrm{14-195 keV}},
\end{equation} 
where $L_{\mathrm{AGN}}$ is the bolometric luminosity of the AGN and $L_{\mathrm{14-195 keV}}$ is the \emph{Swift} BAT luminosity in the 14-195 keV band.

For the ULIRGs, we adopt the starburst and AGN luminosities from V13, which were calculated as follows: the bolometric luminosities were estimated to be $L_{\rm BOL}$ = 1.15 $L_{IR}$, where $L_{\rm IR}$ is the infrared luminosity over 8 -- 1000 $\mu$m \citep{Sanders1996}, and $L_{\rm BOL} = 7 L$(5100~\AA)$ + L_{\rm IR}$ for the PG~QSOs \citep{Netzer2007}. Here, $ L$(5100~\AA) corresponds to $\lambda L_\lambda$ at 5100~\AA.

The starburst and AGN luminosities were next calculated from
\begin{eqnarray}
L_{\rm BOL} & = & L_{\rm AGN} + L_{\rm SB} \\
           & = & \alpha_{\rm AGN}~L_{\rm  BOL} + L_{\rm SB},
\end{eqnarray}
where $\alpha_{\rm AGN}$ is the fractional contribution of the AGN to the bolometric luminosity, hereafter called the ``AGN fraction'' for short. 
For the BAT AGN sample, we derive the AGN fractions from the rest frame $f_{30\,\mu m}/f_{15\,\mu m}$ continuum flux density ratio, which was found by \citet{Veilleux2009} to be more tightly correlated with the PAH-free, silicate-free MIR/FIR ratio and the AGN contribution to the bolometric luminosity than any other {\em Spitzer}-derived continuum ratio. The fraction of the 15~$\mu m$ flux produced by the AGN is defined as:
\input{tab3}
\begin{equation} \label{eq:ne2Frac}
\begin{split}
\frac{\mathrm{AGN}\%(f_{15})}{100}&\equiv\frac{\left(f_{30}/f_{15}\right)_{agn}}{\left(f_{30}/f_{15}\right)_{agn} + \left(f_{30}/f_{15}\right)_{sb}} \\
&=\frac{\left(f_{30}/f_{15}\right)_{obs}-\left(f_{30}/f_{15}\right)_{sb}}{\left(f_{30}/f_{15}\right)_{agn} - \left(f_{30}/f_{15}\right)_{sb}},
\end{split}
\end{equation} 
where $\left(f_{30}/f_{15}\right)_{obs}$ is the observed flux density ratio. $\left(f_{30}/f_{15}\right)_{agn} \mathrm{ and } \left(f_{30}/f_{15}\right)_{sb}$ are the flux density ratios due to the AGN and starburst, respectively. We adopt from Table 9 of \citet{Veilleux2009} the zero-point values of
$\log\left(f_{30}/f_{15}\right)_{agn}=0.2$ and $\log\left(f_{30}/f_{15}\right)_{sb}=1.35$.

The fraction of the bolometric luminosity produced by the AGN is then calculated from:
\begin{equation}
\begin{split}
&\alpha_{\mathrm{AGN}} \equiv \frac{\mathrm{AGN}(L_{\mathrm{bol}})}{100} = \frac{L(bol)^{agn}}{L(bol)^{agn}+L(bol)^{sb}} \\
&=\frac{1}{1+\left[\frac{\mathlarger{100}}{\mathlarger{\mathrm{AGN}\%\left(f_{15}\right)}}-1\right]\left[\frac{\left(\mathlarger{L_\nu\,(15\,\mu \mathrm{m})/L(bol)}\right)_{\mathlarger{agn}}}{\left(\mathlarger{L_\nu\,(15\,\mu \mathrm{m})/L(bol)}\right)_{\mathlarger{sb}}}\right]}
\end{split}
\end{equation}
where we adopt the bolometric corrections of
\begin{align*} 
\log\left[L_\nu\,(15\,\mu \mathrm{m})/L(bol)\right]_{agn}&=-14.33 \\
\log\left[L_\nu\,(15\,\mu \mathrm{m})/L(bol)\right]_{sb}&=-14.56 
\end{align*}
from Table 10 of \citet{Veilleux2009}. As a check, we compare the AGN fractions derived via this method with those found in \citet{Tommasin2010}. For the few objects in common (i.e. NGC 1125, NGC 3516, NGC 4388, NGC 7172, and NGC 7582), we find good agreement between the two methods. For the ULIRGs and QSOs, we adopt the AGN fractions from V13 which are calculated using the same method applied to the BAT AGN.

BAT AGN stellar masses are adopted from \citet{Koss2011}. ULIRG stellar masses are calculated by adopting $H$-band absolute magnitudes from V13. We then assume $M_H^*=-23.7$ (the $H$-band absolute magnitude of a $L^*$ galaxy in a Schechter function description of the local field galaxy luminosity function \citep{Bell2001,Cole2001}) and $m_*=1.4\times10^{11}$ M$_\odot$ (the mass of an early-type galaxy at the knee of the Schechter distribution \citep{Cole2001,Veilleux2002}).

BAT AGN SFRs are derived from equation [25] of \citet{Calzetti2010}:
\begin{equation}
\text{SFR}(160)(M_\odot~\mathrm{ yr}^{-1})=\frac{L(160)~[\text{erg~s}^{-1}]}{4.8\times10^{42}},
\end{equation} 
where $L$(160) is $\lambda L_\lambda$ at 160 $\mu$m and the factor $4.8\times10^{42}$ assumes a \citet{Kennicutt1998} calibration for SFR. SFRs for ULIRGs and PG QSOs are calculated from equation [1] of \citet{RupkeFirst2005}:
\begin{equation}
\text{SFR}=\frac{L_{\mathrm{SB}}}{5.8\times10^9\,L_\odot},
\end{equation}
where $L_{\mathrm{SB}}=L_{\mathrm{BOL}}(1-\alpha_{\mathrm{AGN}})$.

\autoref{fig:distributions} shows the distributions of redshifts, stellar masses, SFRs, AGN fractions, and AGN luminosities for all 52 BAT AGN in our sample (blue diagonals) and for all 43 objects (38 ULIRGs + 5 PG QSOs; red diagonals) from V13. The AGN fractions of the BAT AGN are typically higher than the fractions of the V13 sample of objects.The AGN luminosities and SFRs of the BAT AGN, however, are typically lower by two orders of magnitude than those of the ULIRGs and QSOs.

\section{OBSERVATIONS, DATA REDUCTION, AND SPECTRAL ANALYSIS}
\subsection{OH 119 $\mu$m Feature}
\subsubsection{OH Observations}
The OH observations were obtained with the PACS FIR spectrometer \citep{Poglitsch2010} on board \emph{Herschel} \citep{Pilbratt2010}. We focus our efforts on the ground-state OH 119 $\mu$m $^2\Pi_{3/2}\, J=5/2-3/2$ rotational $\Lambda$-doublet. This
\begin{figure*}[ht!]
\label{fig:ohEqwMassFracLumO4Ne2}
\centering
\subfloat{\includegraphics[scale=0.3]{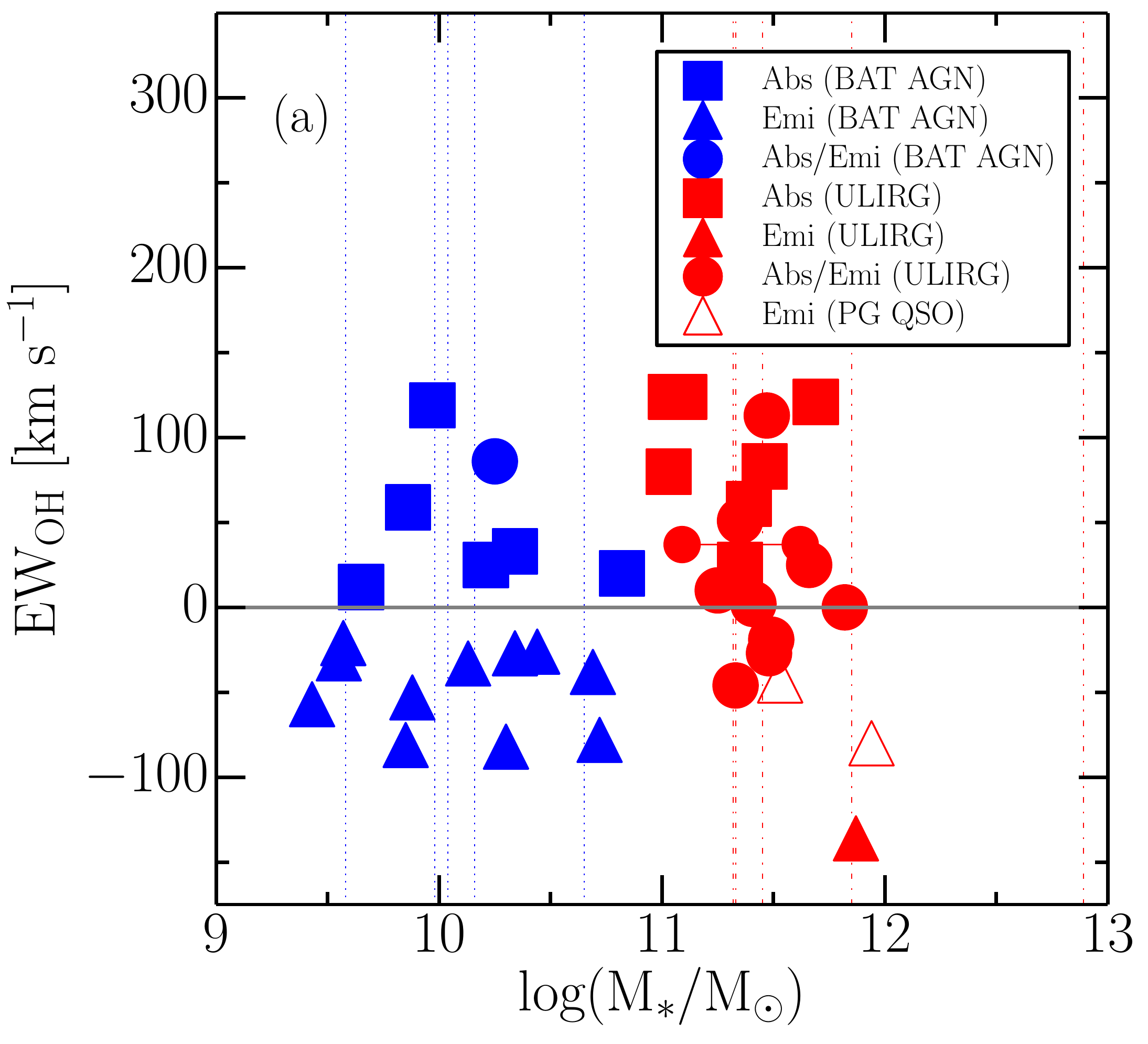}}
\subfloat{\includegraphics[scale=0.3]{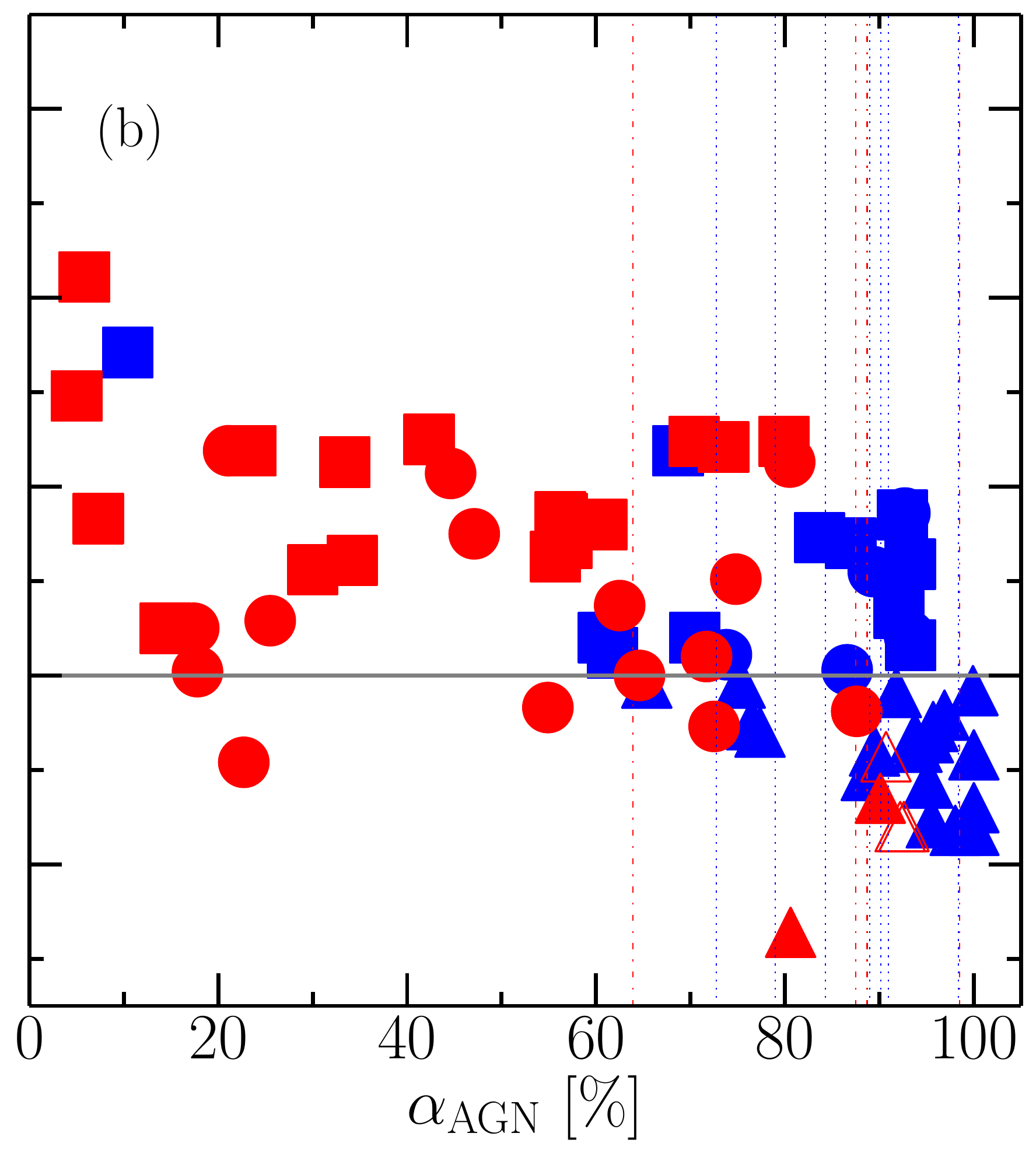}}
\subfloat{\includegraphics[scale=0.3]{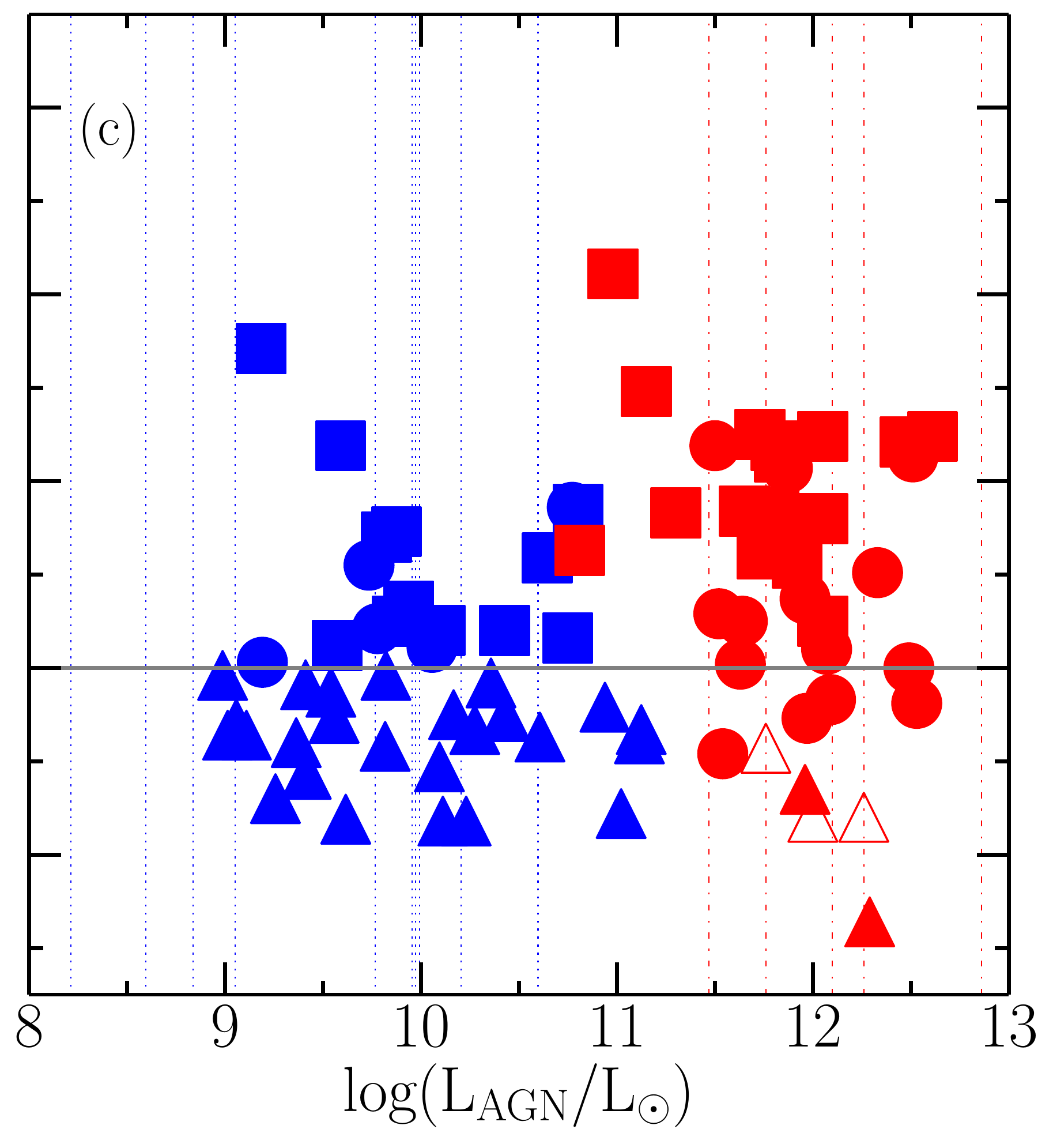}}
\caption{Total (absorption + emission) equivalent widths of OH 119 $\mu$m (positive values indicate absorption and negative values indicate emission) as a function of the (a) stellar masses, (b) AGN fractions, (c) AGN luminosities. The colors blue and red refer to BAT AGN and ULIRGs/PG QSOs, respectively. Filled squares, triangles, and circles represent BAT AGN or ULIRGs in which OH 119 $\mu$m is seen purely in absorption, purely in emission, or with composite absorption/emission, respectively. Open triangles represent PG QSOs in which OH 119 $\mu$m is seen purely in emission. The blue, vertical, dotted lines refer to BAT AGN in which OH 119 $\mu$m is not detected. Similarly, the red, vertical, dash-dotted lines represent ULIRGs/PG QSOs with undetected OH 119 $\mu$m. The gray, horizontal line marks the null OH equivalent width.  }
\end{figure*}
\noindent feature is the strongest transition in ULIRGs and is positioned near the peak spectroscopic sensitivity of PACS. \citet{Fischer2010}, S11, and V13 have demonstrated the efficacy of using the OH 119 $\mu$m feature to determine wind characteristics.

The data for OT2\_sveilleu\_6 were obtained in a similar fashion as those in SHINING (S11) and OT1\_sveilleu\_1 (V13). PACS was used in range scan spectroscopy mode in high sampling centered on the redshifted OH 119 $\mu$m + $^{18}$OH 120 $\mu$m complex with a velocity range of $\sim \pm$4000 km s$^{-1}$ (rest-frame 118-121 $\mu$m) to provide enough coverage on both sides of the OH complex for reliable continuum placement. As a result, the PACS spectral resolution is $\sim$ 270 km s$^{-1}$. The total amount of observation time (including overheads) for OT2\_sveilleux\_6 was 35.3 hours. The program ID and observing time for each target, including all overheads, are listed in \autoref{tab:batAgnProperties}. A large chopper throw of 3$'$ is used in all cases. 

\subsubsection{OH Data Reduction}
 The reduction of the OH data on the BAT AGN sample was carried out using the standard PACS reduction and calibration pipeline (ipipe) included in HIPE 6.0. For the final calibration, fainter sources were normalized to the telescope flux (which dominates the total signal) and recalibrated using a reference telescope spectrum obtained from dedicated Neptune observations during the \emph{Herschel} performance verification phase. \citet{Sturm2011} demonstrated that this telescope background technique can reliably recover the continuum from faint sources. In the following, we use the spectrum of the central $9.4\arcsec \times 9.4\arcsec$ spatial pixel (spaxel) only, without the application of a point source flux correction. In a few objects we note that the OH emission is extended. These objects will be the focus of a future paper. 

The reduced spectra were next smoothed using a Gaussian kernel of width 0.05 $\mu$m (i.e. about half a resolution element) to reduce the noise in the data before the spectral analysis. A spline was fit to the continuum emission and subtracted from the spectra. Subsequently, spectral fitting was carried out on these continuum-subtracted spectra. 

\subsubsection{Spectral Analysis of the OH Doublet} \label{sec:ohAnalysis}
Line profile fits of the OH 119.233, 119.441 $\mu$m doublet were computed by using \emph{PySpecKit}, a spectroscopic analysis and reduction toolkit for optical, infrared, and radio spectra \citep{Ginsburg2011}. The toolkit uses the Levenberg-Marquardt technique to solve the least-squares problem in order to find the best fit for the observations.  Profile fitting of the OH doublet followed a similar procedure as that outlined in V13, in which the doublet profile was modeled using four Gaussian components (two components for each line of the doublet), each characterized by their amplitude (either negative or positive), peak position, and standard deviation (or, equivalently FWHM). However, many of the OH profiles observed here were fit with only two Gaussian components (one component for each line of the doublet), since a fit with four Gaussian components often led to spurious results (i.e. skinny components with FWHM $\sim$ 10 km s$^{-1}$, values too small to be considered real). The separation between the two lines of the doublet was set to 0.208 $\mu$m in the rest frame ($\sim$ 520 km s$^{-1}$) and the amplitude and standard deviation were fixed to be the same for each component in the doublet. In cases where OH was not detected, two Gaussian components, characterized by an amplitude consistent with the 1$\sigma$ level of the noise and a FWHM  (300 km s$^{-1}$) approximately equal to the resolution of PACS, were fit and maximum values for the OH flux and equivalent width were derived. 

Four distinct scenarios apply to our data: (1) pure OH absorption, (2) pure OH emission, (3) P Cygni profiles, and (4) inverse P Cygni profiles. In scenario 1, there is no evidence of OH emission and each line of the OH doublet is fitted with 1-2 absorption components. In the case of the two-component fit, one component traces the stronger low-velocity component of the outflow, while the other component traces the fainter high-velocity component. In the case of the single-component fit, only the low-velocity component is captured. Scenario 2 is treated similarly. In this scenario, there is no evidence for any OH absorption and 1-2 Gaussian components are used to model each line of the OH doublet. In scenario 3, each line of the doublet is modeled with a single blueshifted absorption and a single redshifted emission component. In scenario 4, each line of the doublet is modeled with a single blueshifted emission component and a single redshifted absorption component. In scenarios 3 and 4, just as in the one-component fits of Scenarios 1 and 2, only the low-velocity component of the outflow is captured. 

These fits were first used to quantify the strength and nature (absorption versus emission) of the OH feature: (1) the total flux and equivalent width of the OH 119.441 $\mu$m line, adding up all of the absorption and emission components, (2) the flux and equivalent width of the absorption component(s) used to fit this line, and (3) the flux and equivalent width of the emission component(s) used to fit this line.

Following the same method as that outlined in V13, we also characterize the OH profile by measuring velocities: (1) $v_{50}$(abs) is the median velocity of the fitted absorption profile, i.e., 50\% of the absorption takes place at velocities above (more positive than) $v_{50}$(abs), (2) $v_{84}$(abs) is the velocity above which 84\% of the absorption takes place, (3) $v_{50}$(emi) is the median velocity of the fitted emission profile, i.e., 50\% of the emission takes place at velocities below (less positive than) $v_{50}$(emi), and (4) $v_{84}$(emi) is the velocity below which 84\% of the emission takes place. We note that three objects are fitted with inverted P-Cygni profiles, suggesting inflow. Circinus is an unambiguous fit while Centaurus~A and NGC 3281 are marginally fit with inverted P-Cygni profiles. The velocities from the inverted P-Cygni profile fits are measured as thus: $v_{50}$(abs) and $v_{84}$(abs) are the velocities \emph{below} which 50\% and 84\% of the absorption takes place, respectively, and $v_{50}$(emi) and $v_{84}$(emi) are the velocities \emph{above} which 50\% and 84\% of the emission takes place, respectively. 

We note that the OH emission is extended in some of our BAT AGN sources. While a full analysis of the 5 $\times$ 5 spaxels spectrum is outside the scope of this paper, we provide the ratio of the central spaxel 119 $\mu$m continuum flux density to that of the summed 25 spaxels ($f_{cen}/f_{tot}$) to quantify the degree to which some of our BAT AGN are spatially extended (see \autoref{tab:batAgnProperties}). For reference, the average value of $f_{cen}/f_{tot}$ for a point source (derived from the five PG QSOs in our sample) is 0.56. Note that the continuum is often considerably more extended than the OH 119 $\mu$m feature. 

\begin{figure}
\centering
\label{fig:agnFracS97V2}
\includegraphics[scale=0.4]{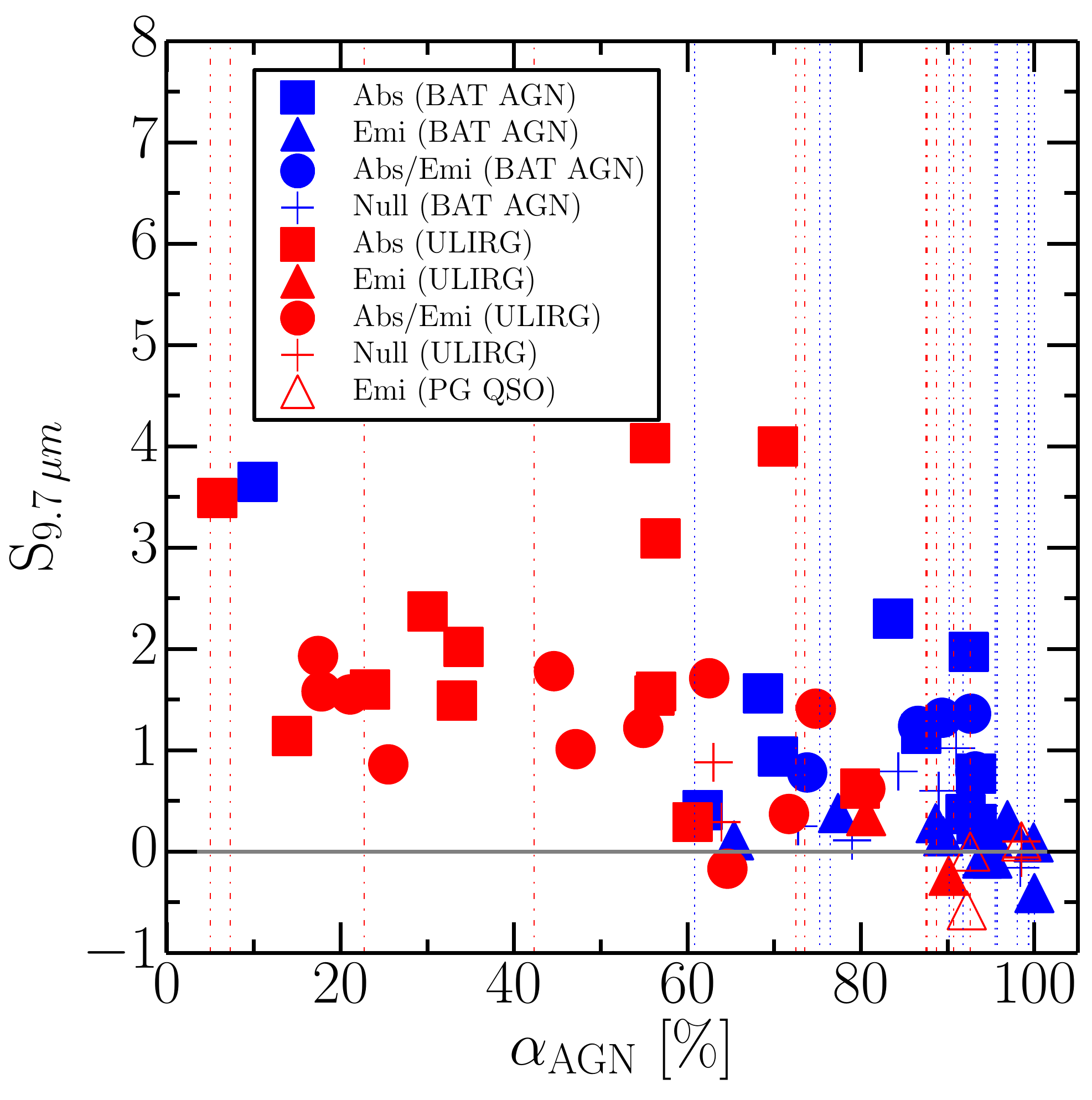}
\caption{The apparent strength of the 9.7 $\mu$m silicate feature relative to the local mid-infrared continuum as a function of the AGN fractions. Note that \silicate is a logarithmic quantity and can be interpreted as the apparent silicate optical depth. The strength of silicate absorption increases upward. Sign conventions and meanings of the symbols are the same as those in \autoref{fig:ohEqwMassFracLumO4Ne2}. Crosses represent objects with a null OH detection. Vertical lines indicate objects with a null 9.7 $\mu$m silicate feature detection.}
\end{figure}
\subsection{The 9.7 $\mu$m Silicate Feature}
\subsubsection{Data Reduction of the 9.7 $\mu$m Silicate Feature}
Mid-infrared (MIR; 5-37 $\mu$m) low resolution spectra (R $\sim 60-127$) used to measure the 9.7 $\mu$m silicate feature for the BAT AGN, ULIRG, and PG QSO samples were extracted from The Cornell AtlaS of Spitzer/IRS Sources (CASSIS\footnotemark[1]).
\footnotetext[1]{http://www.cassis.sirtf.com/atlas}
 The archival data were obtained with the Infrared Spectrograph (IRS; \cite{Houck2004}) on board the Spitzer Space Telescope \citep{Werner2004}. There were two objects however (NGC 7479 and NGC 4945), in which the MIR spectrum was extracted from the Spitzer Heritage Archive (SHA\footnotemark[2]).
 \footnotetext[2]{http://sha.ipac.caltech.edu/applications/Spitzer/SHA/}

 The \emph{Spitzer} observations were made with the Short-Low (SL, 5.2-14.5 $\mu$m) and Long-Low (LL, 14.0-38.0 $\mu$m) modules of the IRS. The orders were stitched to LL order 2, requiring order-to-order scaling adjustments of less than $\sim$ 15\%.

\subsubsection{Spectral Analysis of the 9.7 $\mu$m Silicate Feature}\label{sec:s97Analysis}

We have measured the strength of the 9.7 $\mu$m silicate feature for sources in our BAT AGN sample and in the ULIRG/QSO sample. The calculation of the mid-infrared continuum loosely follows the method of \citet{Spoon2007}. For a majority of the sources, the mid-infrared continuum (e.g. the continuum flux density $f_{cont}$) is determined from a cubic spline interpolation to continuum pivots at 5.2, 5.6, 14.0, 27.0, and 31.5 $\mu$m. For objects in which the 9.7 $\mu$m silicate feature dominates the spectrum (i.e. there is very little PAH emission), an additional pivot point is added at 7.8 $\mu$m. The pivot points are adopted from \citet{Spoon2007}, however we have placed a pivot point at 27 $\mu$m instead of at 25 $\mu$m due to the proximity of the [OIV] emission line at 25.89 $\mu$m which is common in BAT AGN. Due to the diversity of our BAT AGN sample, the wavelength range in which the 9.7 $\mu$m silicate feature is determined differs slightly from source to source (see \autoref{tab:batSilicateProperties}). Typically, however, the observed flux density (e.g. $f_{obs}$) is determined from a cubic spline interpolation to the data between $\sim 8 \mu$m  and $\sim 14 \mu$m. This interpolation skips the H$_2$ line at 9.6 $\mu$m, [S IV] line at 10.51 $\mu$m, PAH feature at 11.25 $\mu$m, H$_2$ line at 12.28 $\mu$m, and [NeII] at 12.68 $\mu$m.

The apparent strength of the 9.7 $\mu$m silicate feature is then defined as:
\begin{equation}\label{eq:s97equation}
S_{9.7\,\mu\mathrm{m}}=-\ln\left(\frac{f_{obs}(9.7\,\mu\mathrm{m})}{f_{cont}(9.7\,\mu\mathrm{m})}\right).
\end{equation}

Here, we adopt the sign convention consistent with our OH 119 $\mu$m analysis (e.g. positive \silicate values indicate absorption while negative \silicate values indicate emission). Note that this convention is different from previous studies \citep[e.g.][]{Spoon2007}.

For sources with a silicate absorption feature, \silicate can be interpreted as the the apparent silicate optical depth. \autoref{tab:batSilicateProperties} and \autoref{tab:ulirgSilicateProperties} lists pivot points, integration ranges, and measured equivalent widths and fluxes of the 9.7 $\mu$m silicate feature for our samples. MIR spectra showing these pivot points and integration ranges may be found in the Appendix.

\begin{figure}
\centering
\label{fig:ohS97V2}
\includegraphics[width=\columnwidth]{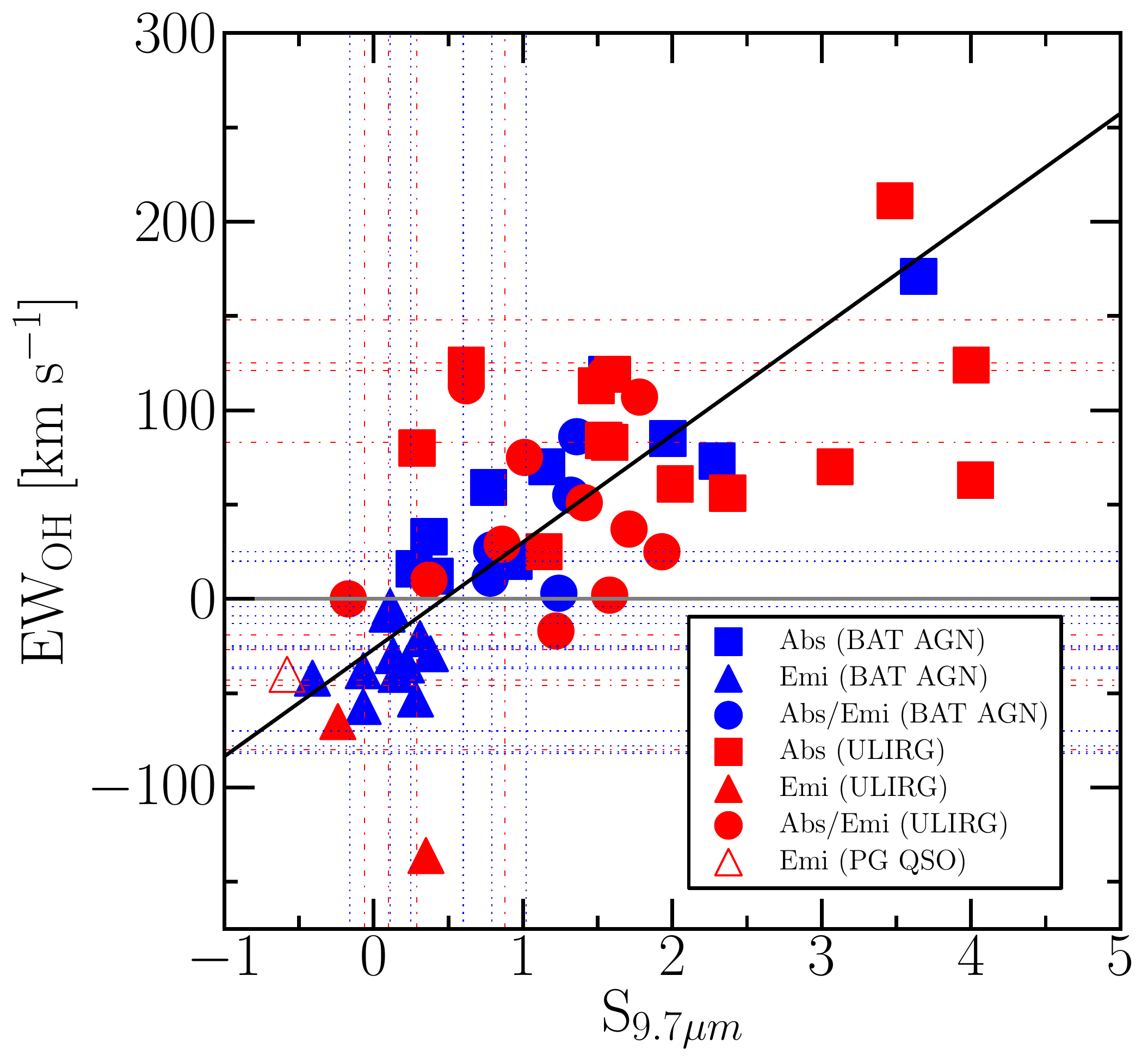}
\caption{Total (absorption + emission) equivalent widths of OH 119 $\mu$m as a function of the apparent strength of the 9.7 $\mu$m silicate feature relative to the local mid-infrared continuum. Note that \silicate is a logarithmic quantity and can be interpreted as the apparent silicate optical depth. The strength of this absorption feature increases to the right. Sign conventions and meanings of the symbols are the same as those in \autoref{fig:ohEqwMassFracLumO4Ne2}. Horizontal, blue, dotted lines represent BAT AGN in which the 9.7 $\mu$m silicate feature is not detected. Similarly, horizontal, red, dash-dotted lines represent ULIRGs/PG QSOs with a null 9.7 $\mu$m silicate feature detection. The black diagonal line shows the ordinary least squares bisector linear regression: EW$_{\mathrm{OH}} = 56.81\times $\silicate$-26.7~~{\rm km s}^{-1}$.  The Pearson r null probability for the linear relationship between \silicate and the total OH equivalent width is $P$[null]$ = 1.2\times10^{-9}$ for the BAT AGN sample and $P$[null] = 0.003 for the ULIRG + QSO sample. When the samples are combined, we find $P$[null] = $3.9\times10^{-9}$. }
\end{figure}

\section{Results}

\autoref{fig:ohSpectralFits} shows the fits to the OH 119 $\mu$m profiles for our BAT AGN targets. The OH 119 $\mu$m parameters for all targets derived from these fits are listed in \autoref{tab:ohProperties}. The meaning of each parameter is discussed in \autoref{sec:ohAnalysis} and in the notes to \autoref{tab:ohProperties}. Note that the fluxes and equivalent widths in \autoref{tab:ohProperties} need to be multiplied by a factor of two when considering both lines of the doublet. 

In this section we compare the OH 119 $\mu$m and \silicate results listed in \autoref{tab:ohProperties}, \autoref{tab:batSilicateProperties}, and \autoref{tab:ulirgSilicateProperties}, with the galaxy properties listed in \autoref{tab:batAgnProperties}.

\begin{figure*}
\centering
\label{fig:velNonDistMultiHistV4}
\includegraphics[scale=0.5]{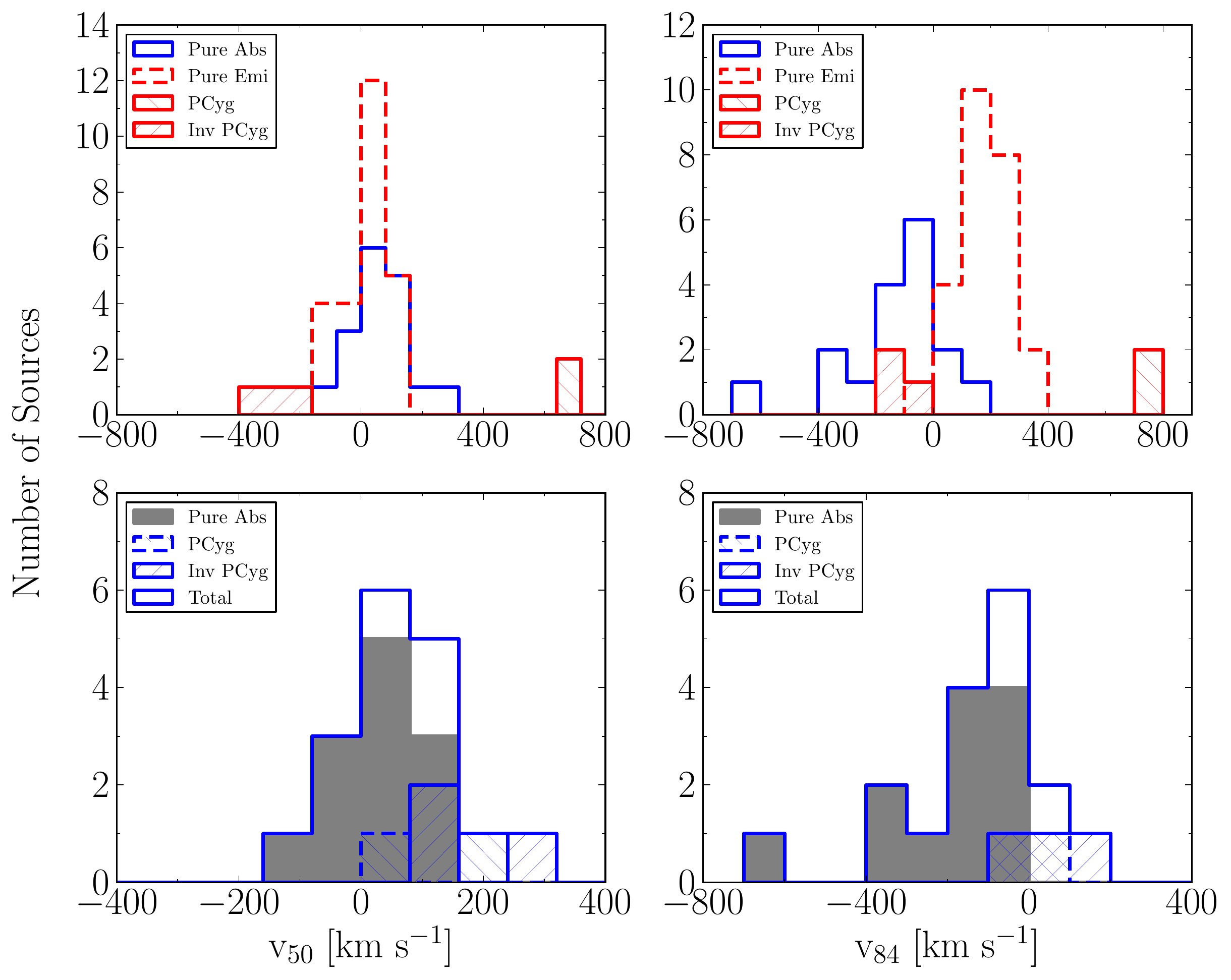}
\caption{Histograms showing the distributions of the 50\% (median; left panels) and 84\% (right panels) velocities derived from the multi-Gaussian fits to the OH profiles of the BAT AGN. Top panels show pure absorption components (blue), pure emission components (red dashed), P-cygni emission components (red, left diagonals), and inverse P-cygni emission components (red, right diagonals). Bottom panels show pure absorption components (filled grey), P-Cygni absorption components (blue, left diagonals), inverse P-cygni absorption components (blue, right diagonals), and total absorption components (pure + P-Cygni + inverse P-Cygni).}
\end{figure*}
\subsection{The OH 119 $\mu$m Feature}

The OH 119 $\mu$m doublet is detected in 42 of the 52 objects in our BAT AGN sample. Of the 42 detections, 25 are seen purely in emission, 12 are seen purely in absorption, and 5 are seen with absorption+emission composite profiles. For comparison, in the ULIRG + QSO sample (V13), 37 objects showed OH 119 $\mu$m detections, 17 of which were seen purely in absorption, 15 showed absorption+emission composite profiles, and 5 were seen purely in emission. \autoref{fig:ohEqwMassFracLumO4Ne2}a, \autoref{fig:ohEqwMassFracLumO4Ne2}c plot the OH equivalent width with the stellar mass and AGN luminosity, respectively. For these properties, we see no discernible correlation in either the individual samples (i.e. BAT AGN only and ULIRG + PG QSO only) or in the combined sample. There is however a weak trend between EW$_\mathrm{OH}$ and AGN fraction (\autoref{fig:ohEqwMassFracLumO4Ne2}b) when the samples are combined. A more formal statistical analysis of these parameters indeed indicates a statistically significant correlation between these quantities when all objects are considered (see \autoref{tab:statResultsHostGalaxy}).

For the BAT AGN in which OH 119 $\mu$m is seen purely in emission, we find that the AGN is dominant (i.e. $\alpha_{\mathrm{AGN}} > 50\%$) in these objects. In agreement with V13, the low luminosities ($\log (L_{\mathrm{AGN}}/L_\odot) \lesssim 12$) of these 20 AGN-dominated objects with pure OH emission suggest that the AGN fraction is more important than AGN luminosity in setting the character (i.e. strength of emission relative to absorption) of the OH feature. Therefore, we see that dominant AGN in our sample can excite the OH molecule (via radiative pumping or collisional excitation) and produce the $^2\Pi_{3/2}\,J = 5/2\rightarrow3/2$ rotational emission line. This trend was noted in V13. 

\input{tab4}

\subsection{The \silicate Feature}
\autoref{fig:agnFracS97V2} plots the apparent strength of the 9.7 $\mu$m silicate absorption feature versus the AGN fraction $\alpha_{\mathrm{AGN}}$. Here, larger, more positive values of \silicate indicate stronger silicate absorption features. We see no discernible trend between the strength of this silicate feature and AGN fraction. This is surprising since the AGN fractions should be inversely proportional to the equivalent widths of the PAH features \citep{Veilleux2009}. Thus, \autoref{fig:agnFracS97V2} should look similar to Figure 1 of \citet{Spoon2007} which plots the 6.2 $\mu$m PAH emission feature and \silicate. We find that our data do not show the bifurcation observed in \citet{Spoon2007} because only two of our ULIRGs (F08572+3915 and F15250+3608) lie on the upper branch of the bifurcation, and none of our BAT AGN populate the upper branch. The two objects in \autoref{fig:agnFracS97V2} which display $\alpha_\mathrm{AGN}\sim10\%$ and high \silicate are NGC 4945 and F15327+2340. Both of these objects show weak PAH emission and are therefore not displayed in Figure 1 of \citet{Spoon2007}.

\autoref{fig:ohS97V2} shows a positive correlation between the measured OH equivalent width and the 9.7 $\mu$m silicate optical depth. The black diagonal line shows the ordinary least squares bisector linear regression for the entire sample (BAT AGN + ULIRGs + PG QSOs):
\begin{equation}\label{eq:bisector}
\mathrm{EW}_{\mathrm{OH}}= 56.81\times S_{9.7\,\mu\mathrm{m}}-26.7~[\mathrm{km~ s}^{-1}].
\end{equation}
The Pearson correlation coefficient of the linear relationship for \silicate and the total OH equivalent width is $\rho_{S_{9.7\,\mu m},\,\mathrm{EW}_{\mathrm{OH}}}=0.7$ with a $P$[null] (the probability of an uncorrelated system producing datasets that have a Pearson correlation at least as extreme as the one computed from these datasets) of $3.9\times10^{-9}$ for the combined sample. If we restrict this analysis to the BAT AGN only, we find $\rho_{S_{9.7\,\mu m},\,\mathrm{EW}_{\mathrm{OH}}}(\mathrm{BAT\, AGN})=0.89$ with a $P$[null] of $1.2\times10^{-9}$. Restricting the analysis to ULIRGs and PG QSOs only yields $\rho_{S_{9.7\,\mu m},\,\mathrm{EW}_{\mathrm{OH}}}(\mathrm{ULIRG/PG\, QSO})=0.5$ with a $P$[null] of $0.003$. 

In \autoref{fig:ohS97V2} we also see that objects with OH in emission show either weak silicate absorption or silicate emission. Objects with OH P-Cygni profiles show moderate silicate absorption features, while the strongest silicate absorption features are seen in objects in which OH is observed purely in absorption. We return to this result in \autoref{sec:discussion}. 

\subsection{OH Kinematics}\label{sec:kinematics}

We quantify the visual trends observed (or not observed) in our kinematic investigation of the individual samples (i.e. BAT AGN only or ULIRG/QSO only) and of the combined sample by calculating the correlational significances between the observed OH velocities ($v_{50}$ and $v_{84}$) and the host galaxy properties (stellar masses, SFRs, specific star formation rates (sSFRs; i.e. the rate at which stars are formed divided by the stellar mass of the galaxy), and AGN fractions and luminosities. Since the spatial location of the OH emission is unknown, physical interpretations of the observed OH velocities in these cases will be ambiguous (e.g., blueshifted velocities may correspond to outflow or inflow if the OH emission region is located in front or behind the continuum source, respectively). We therefore exclude from our analysis objects in which OH is detected purely in emission. The results of our statistical analyses on all objects with either redshifted or blueshifted absorption profiles are listed in \autoref{tab:statResults}. 

\autoref{fig:velNonDistMultiHistV4} shows the distributions of velocities derived from both the OH absorption and emission line features ($v_{50}$(abs), $v_{84}$(abs), $v_{50}$(emi), and $v_{84}$(emi), as defined in \autoref{sec:ohAnalysis} and listed in \autoref{tab:ohProperties}). In V13, we adopted \citet{RupkeSecond2005}'s conservative definition of an outflow as having an OH absorption feature with a median velocity ($v_{50}$) more negative than $-$50 km s$^{-1}$. Similarly, we can define an inflow as having an OH absorption feature with a median velocity ($v_{50}$) more positive than 50 km s$^{-1}$.  This cutoff is used to avoid contamination due to systematic errors and measurement errors in wavelength calibration, line fitting (see \autoref{sec:ohAnalysis}), and redshift determination. We find that only two objects in our BAT AGN sample meet this outflow criterion, and just barely: NGC 7172 and NGC 7479 (both have $v_{50}$ = $-$51 km s$^{-1}$). The presence of a significant blue wing in the absorption profile of OH 119 $\mu$m in NGC 7479 with $v_{84} = -658$ km s$^{-1}$ adds considerable support to this interpretation. An extended blue wing with $v_{84} < -300$ km s$^{-1}$ may also be present in the OH absorption profile of IC 5063, NGC 5506, and NGC 7172, although these detections are more tentative than in NGC 7479. In contrast, seven objects have OH absorption features with median velocities larger than +50 km s$^{-1}$: Centaurus~A, Circinus, NGC 1125, NGC 3079, NGC 3281, NGC 4945, and NGC 7582. The clear detection of an inverted P-Cygni profile in Circinus provides unambiguous evidence for the presence of an inflow in this object. The inverted P-Cygni profiles in Centaurus~A and NGC 3281 are much less secure. We return to these objects in \autoref{sec:discussion}. 

\input{tab5}
\input{tab6}
\input{tab7}

\begin{figure}
\centering
\label{fig:stellarMassVelocity}
\includegraphics[scale=0.5]{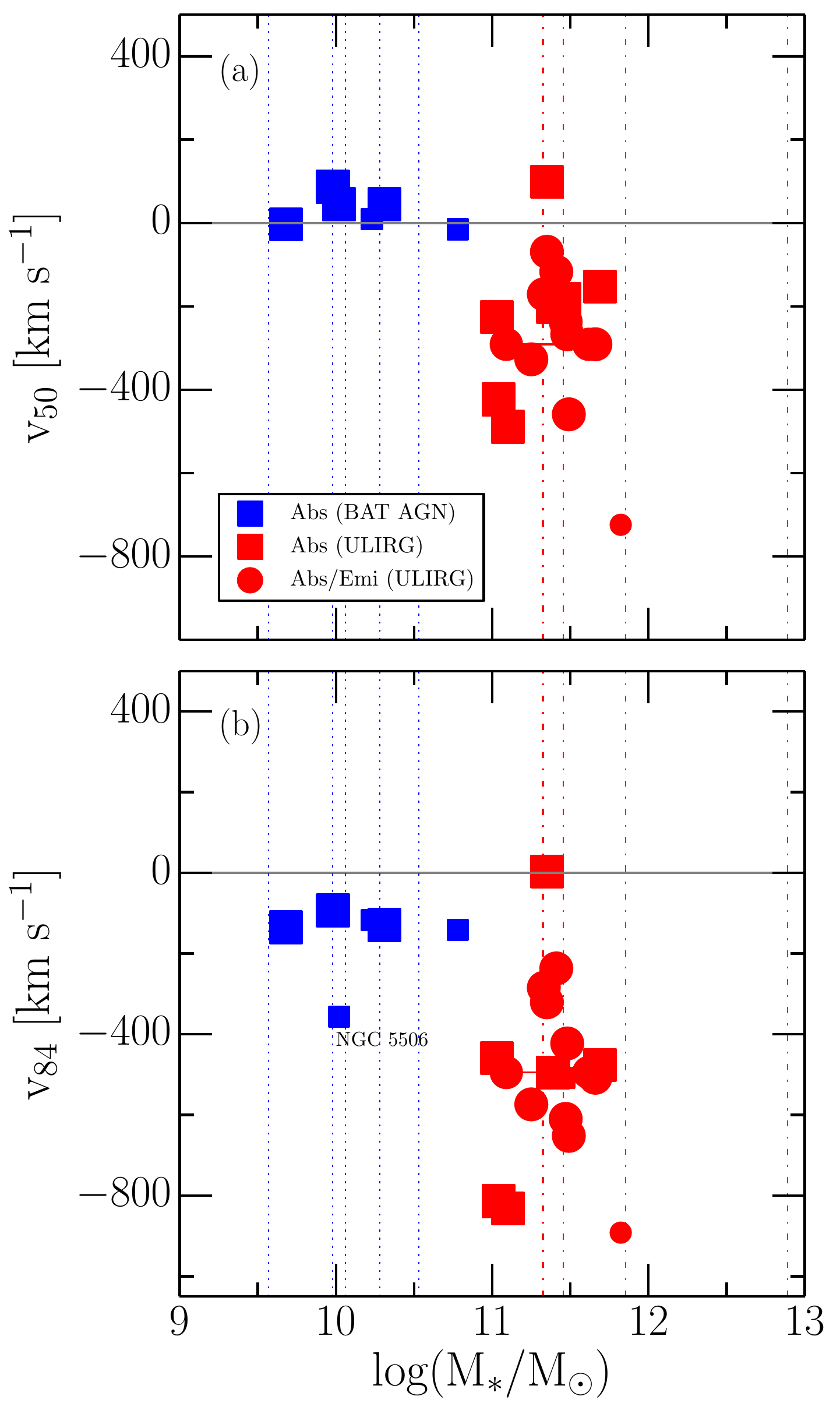}
\caption{$v_{50}$ and $v_{84}$ as a function of the stellar masses. The meanings of the symbols are the same as those in \autoref{fig:ohEqwMassFracLumO4Ne2}. The data points joined by a segment correspond to F14394+5332 W and E. The smaller symbols have larger uncertainties (values followed by double colons in \autoref{tab:ohProperties})}.
\end{figure}

\begin{figure}
\centering
\label{fig:sfrVelocityV2}
\includegraphics[scale=0.5]{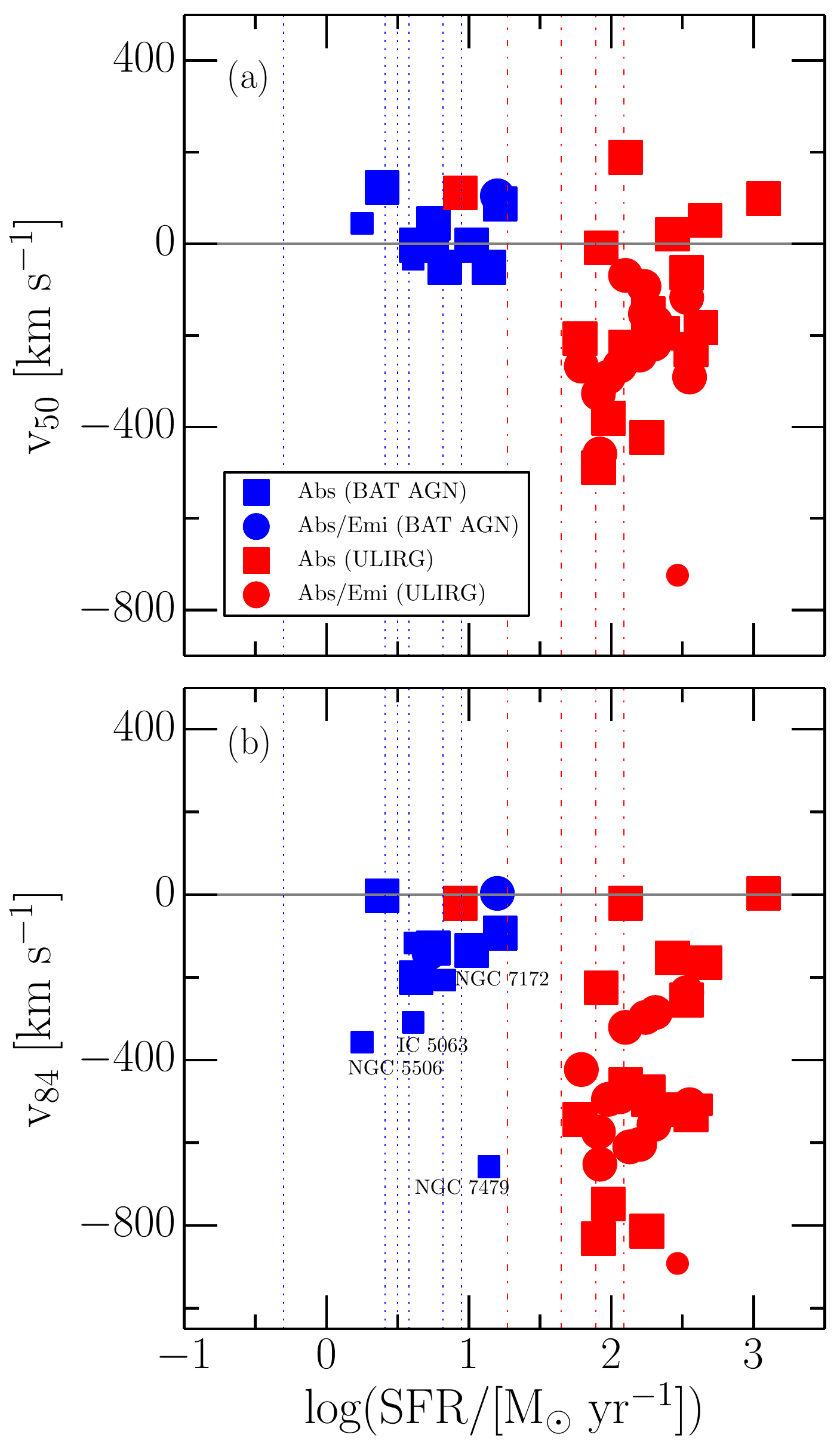}
\caption{$v_{50}$ and $v_{84}$ as a function of the star formation rates. The meanings of the symbols are the same as those in \autoref{fig:ohEqwMassFracLumO4Ne2}. The smaller symbols have larger uncertainties (values followed by double colons in \autoref{tab:ohProperties})}
\end{figure}

\begin{figure}
\centering
\label{fig:sSfrVel}
\includegraphics[scale=0.5]{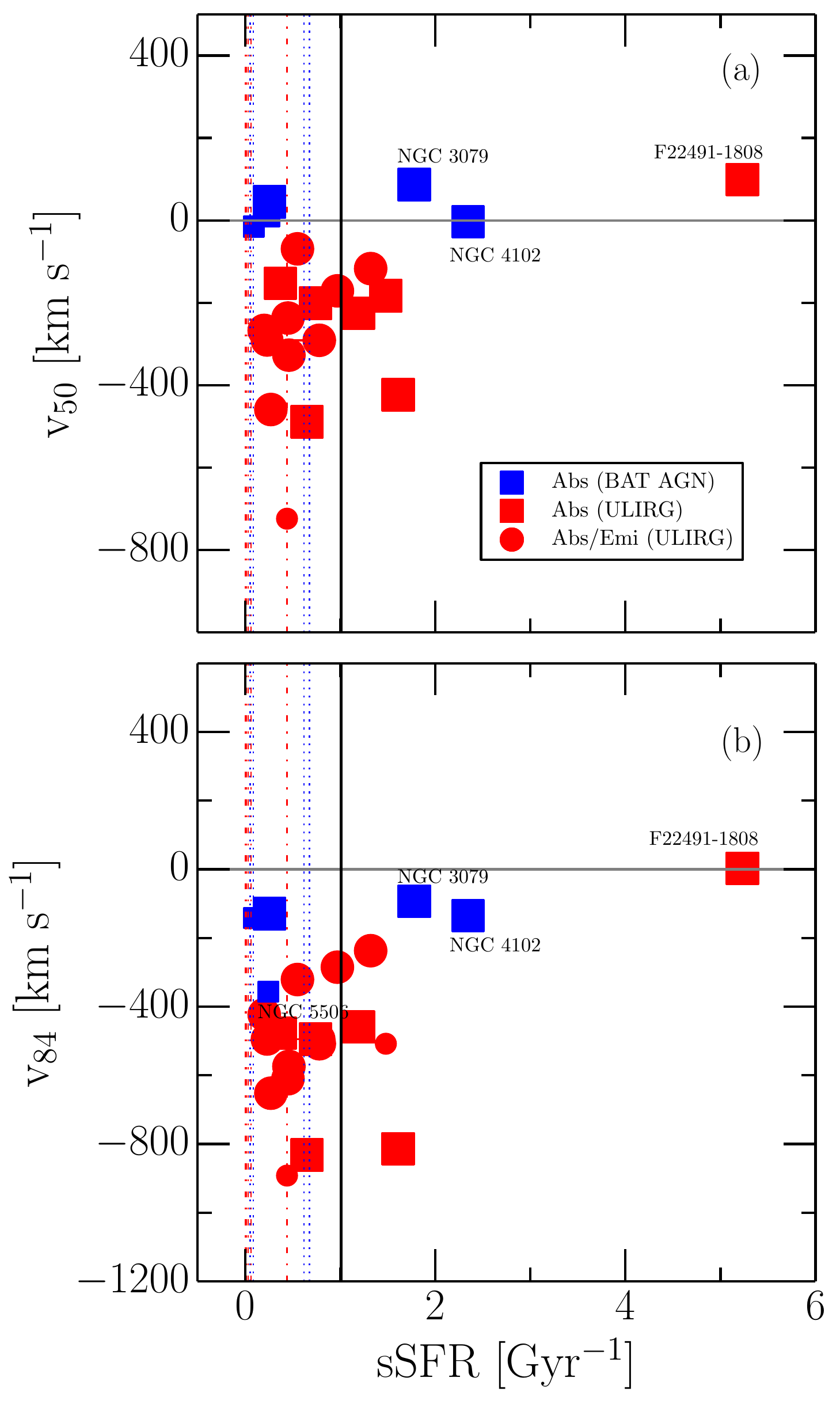}
\caption{$v_{50}$ and $v_{84}$ as a function of the specific star formation rates. The meanings of the symbols are the same as those in \autoref{fig:ohEqwMassFracLumO4Ne2}. The data points joined by a segment correspond to F14394+5332 E and W. The smaller symbols have larger uncertainties (values followed by double colons in \autoref{tab:ohProperties}) The black vertical line indicates the approximate location of the Main Sequence of star-forming galaxies \citep{Shimizu2015}.}
\end{figure}

\autoref{fig:stellarMassVelocity}, \autoref{fig:sfrVelocityV2}, and
\autoref{fig:sSfrVel} plot the OH velocities ($v_{50}$ and $v_{84}$)
as a function of the stellar masses, star formation rates, and
specific star formation rates. If we consider our two samples individually, a visual inspection of these plots does not show a correlation between these properties. However, once the samples are combined we see a negative correlation between the observed OH velocities and the stellar mass or star formation rate of the galaxy
(i.e. galaxies with larger stellar masses or star formation rates
exhibit more negative $v_{50}$ and $v_{84}$ values). The strengths of
these correlations are quantified in \autoref{tab:statResults}. No
obvious trend is seen with sSFR for the combined sample. These results remain quantitatively the same if instead of using the global star formation rates we use
the star formation rates from the central spaxel (scaled by the ratio
of the continuum flux from the central spaxal to that of all 25
spaxels, as listed in column (8) of \autoref{tab:batAgnProperties}).

\begin{figure}
\centering
\label{fig:agnFracMultiPlot}
\includegraphics[scale=0.5]{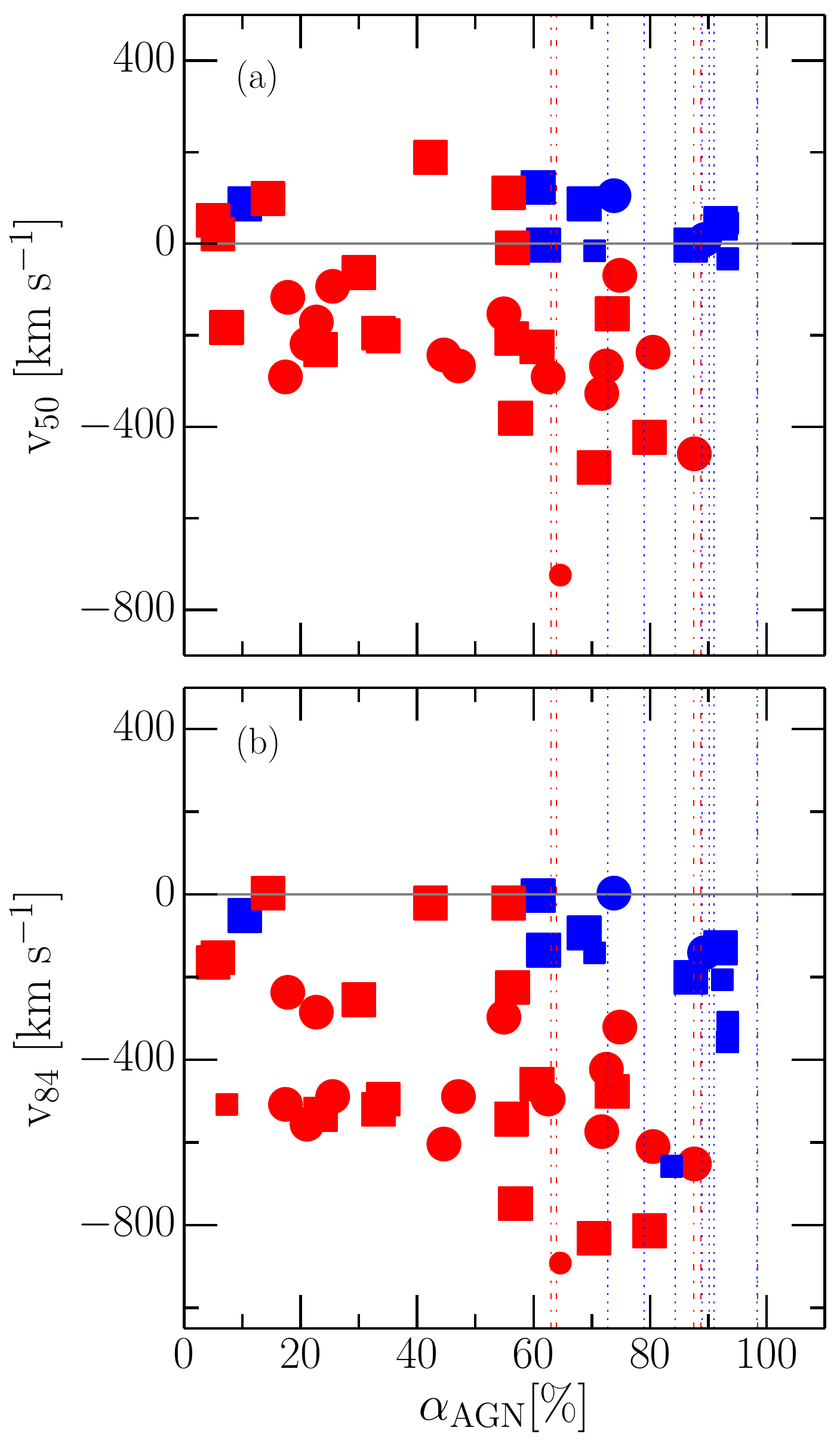}
\caption{$v_{50}$ and $v_{84}$ as a function of the AGN fractions. The meanings of the symbols are the same as those in \autoref{fig:ohEqwMassFracLumO4Ne2}. The smaller symbols have larger uncertainties (values followed by double colons in \autoref{tab:ohProperties})}
\end{figure}

\autoref{fig:agnFracMultiPlot} plots the OH velocities ($v_{50}$ and $v_{84}$) as a function of the AGN fractions $\alpha_{\mathrm{AGN}}$, where $\alpha_{\mathrm{AGN}}$ is derived from the $f_{30}/f_{15}$ continuum flux density ratios. As described in V13, ULIRGs/PG QSOs with dominant AGN ($\alpha_{\mathrm{AGN}} \geq 50\%$) appear to have larger negative velocities than ULIRGs/PG QSOs with dominant starbursts ($\alpha_{\mathrm{AGN}} \leq 50\%$), but a K-S test between velocity distributions of dominant AGN and dominant starburst systems indicated no statistically significant difference. A K-S test on the combined BAT AGN and ULIRG + QSO sample also does not show a significant trend between the measured OH velocity and AGN fraction. 

\begin{figure}
\centering
\label{fig:velLumPlotWithUlirg}
\includegraphics[scale=0.5]{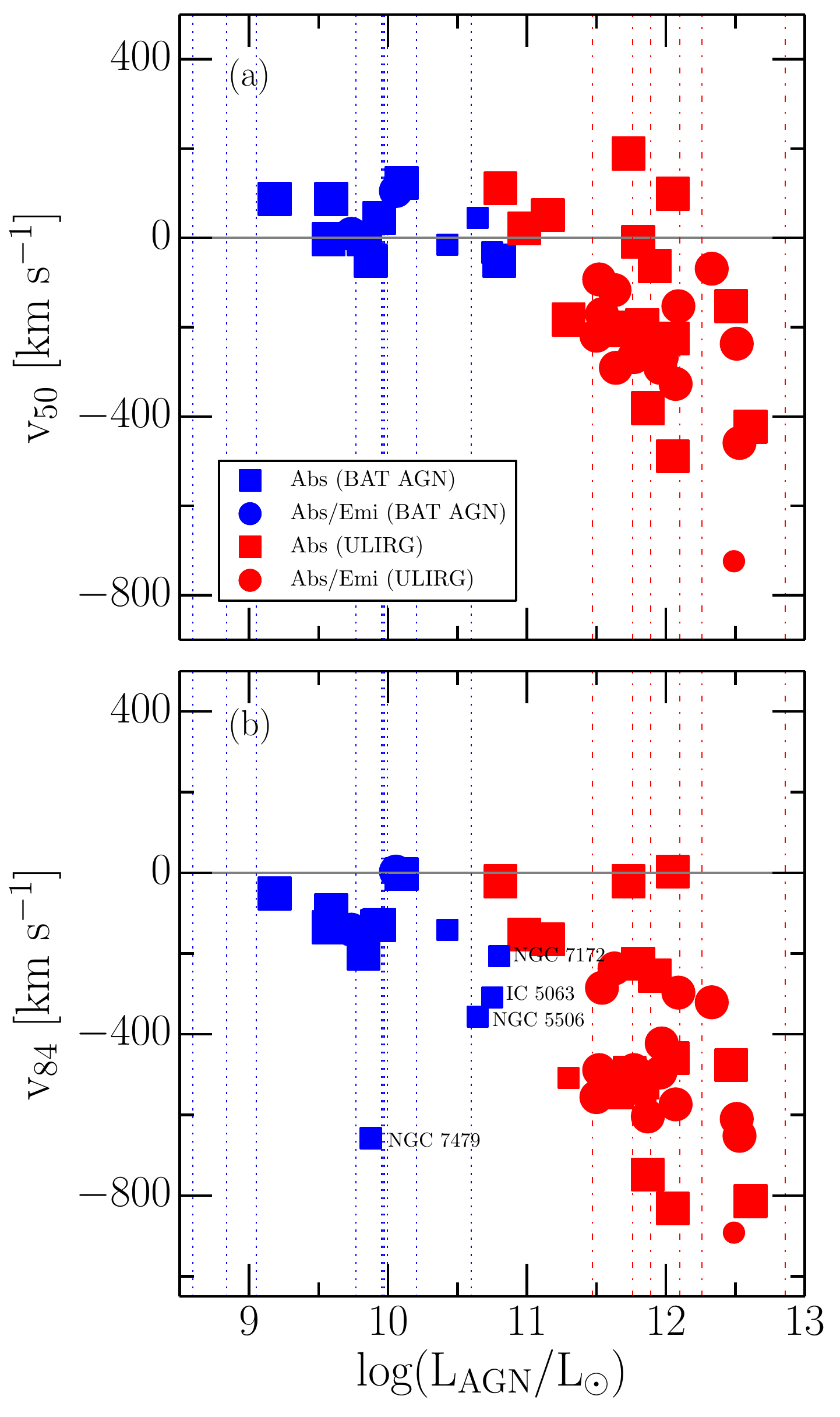}
\caption{$v_{50}$ and $v_{84}$ as a function of the AGN luminosities. The meanings of the symbols are the same as those in \autoref{fig:ohEqwMassFracLumO4Ne2}. The smaller symbols have larger uncertainties (values followed by double colons in \autoref{tab:ohProperties})}
\end{figure}

\autoref{fig:velLumPlotWithUlirg} shows the OH velocities ($v_{50}$ and $v_{84}$) versus the AGN luminosities, $L_{\mathrm{AGN}} = 10.5\times L_{14-195\mathrm{ keV}}$ for our BAT AGN sample and $L_{\mathrm{AGN}}=\alpha_{\mathrm{AGN}}L_{\mathrm{BOL}}$ for the ULIRG/QSO sample; see \autoref{sec:galaxyProperties}. A correlation between these quantities is not observed in the BAT AGN sample, but clear trends are seen in the ULIRG/QSO sample and in the combined sample (see \autoref{tab:statResults}). Objects with $\log\left(L_{\mathrm{AGN}}/L_\odot\right) \lesssim$11.5 show no evidence for fast outflows. This suggests that AGN of lower luminosities are not able to drive significant molecular winds. 

\section{Discussion}\label{sec:discussion}

\subsection{Outflows}

\subsubsection{BAT AGN with Molecular Outflows}\label{sec:batMolOutflows}

As mentioned above, only (one) four objects in our BAT AGN sample shows (unambiguous) evidence of a molecular outflow (NGC 7479 is the unambiguous case, while NGC 5506, NGC 7172, and IC 5063 are more uncertain; see \autoref{fig:sfrVelocityV2} and \autoref{fig:velLumPlotWithUlirg}). This corresponds to an outflow detection rate of (6\%) 24\%, if we take into account that this search for outflows was possible only in the 17 sources with OH 119 $\mu$m in absorption. This outflow detection rate is significantly smaller than that found in ULIRGs ($\sim$70\%). We note, however, that outflow detection may be easier in the ULIRGs due to their higher gas fractions.

NGC 7479 is the best case for an outflow in our sample of BAT AGN. NGC 7479 is a barred galaxy with the faint broad line emission at H$\alpha$ but not at H$\beta$ of a Seyfert 1.9 galaxy \citep{Veron2006}. To quantify the uncertainty of the OH blue-wing detection in NGC 7479, we have refit the continuum for this object with a third order polynomial and then refit the OH doublet via the method outlined in \autoref{sec:ohAnalysis}. Although we find a shift in the velocities measured with this different continuum ($v_{50}$(abs) = $-$45 km s$^{-1}$ and $v_{84}$(abs) = $-$513 km s$^{-1}$), this shift (a measure of the error in our velocity estimates) is not large enough to account for the location of NGC 7479 in \autoref{fig:velLumPlotWithUlirg}b. The origin for the high outflow velocity in NGC 7479 may lie in its unusually high far-infrared surface density \citep{Lutz2015}.  The high wind velocity observed in NGC 7479 may thus be due to its unusually high SFR surface density \citep[e.g.][]{Diamond2012}. However, we cannot exclude the possibility that the OH outflow in NGC 7479 is driven by the radio jet detected by \citet{Laine2008}.

NGC 5506 is an edge-on disk galaxy optically classified as a Seyfert 1.9 galaxy, just like NGC 7479. In the very hard X-ray regime, NGC 5506 is one of the most luminous and brightest Seyferts in the local universe \citep[L$_{14-195\,\mathrm{ keV}} \sim 10^{43}$ erg s$^{-1}$;][]{Baumgartner2011}, and its obscuring column \citep[$N_\mathrm{H} = 3.4 \times 10^{22}$ cm$^{-2}$;][]{Bassani1999} is intermediate between typical values for Seyfert 1s and 2s. Mid-infrared $8-13 \mu$m observations of NGC 5506 \citep{Roche1984} suggest that the nuclear region of NGC 5506 within 5 arcsec is powered by highly obscured AGN activity with little starburst activity. The OH 119 $\mu$m line profile of NGC 5506 shows a very broad, secondary gaussian component, suggestive of an outflow (see \autoref{fig:ohSpectralFits}), but the velocities of this component are uncertain (see \autoref{tab:ohProperties}). Interestingly, the optical forbidden emission lines of NGC 5506 also exhibit distinct blue wings extending to $-$1000 km s$^{-1}$ \citep{Veilleux1991a, Veilleux1991b, Veilleux1991c}, giving credence to the idea that a fast outflow is indeed present in this object. 

IC 5063 is a massive (M$_* \approx10^{11}$ M$_\odot$) early-type galaxy which hosts a powerful double-lobed radio source ($P_{1.4\mathrm{ GHz}}=3\times10^{23}\text{ W Hz}^{-1}$). The presence of this strong radio source is contrary to the weakly collimated jets found in the BAT AGN of \citet{Middelberg2004}. Observations of high velocity ($\sim$~600~km~s$^{-1}$) warm molecular hydrogen gas in the western lobe of IC 5063 suggests that molecules may be shock-accelerated by the expanding radio jets \citep{Tadhunter2014}. Thus, the presence of this powerful jet may explain the additional boost observed in the (uncertain) outflow velocity $v_{84}$. 

NGC 7172 is an obscured \citep[$N_\mathrm{H}\sim10^{23}$
  cm$^{-2}$;][]{Turner1989}, almost edge-on Type 2 Seyfert galaxy
which belongs to the Hickson compact group HCG 90. \citet{Smajic2012}
derive the two-dimensional velocity field and velocity distribution of
the central $4\arcsec\times4\arcsec$ region of NGC 7172 using several
emission lines (i.e. Pa$\alpha$, H$_2$(1-0)S(1), and [Si VI]) and two
CO stellar absorption features. All measurements indicate disk
rotation from east to west with amplitude of at least $\pm$100 km
s$^{-1}$. Comparison of HST F606W imaging \citep{Malkan1998} with
2MASS images in $J-$, $H-$, and $K-$bands \citep{Jarrett2003} shows a
shift of the nucleus by more than 1\arcsec\ going from the visible to
the $J-$, $H-$, and $K-$band. The central spaxel aperture ($9.4\arcsec
\times 9.4\arcsec$) of {\em Herschel} does not have the resolution to
detect such a shift, but if the central spaxel is not centered on the
galaxy nucleus, and if the velocity structure for OH 119 $\mu$m is
comparable to that seen in the near-infrared, then it is possible that
the blueshifted absorption we detect is due to incomplete sampling of
the rotation curve rather than a molecular outflow.

This last object emphasizes the fact that the poor spatial resolution
of {\em Herschel} makes the interpretation of the OH spectra
challenging. We have tried to err on the side of being cautious by
utilizing a centroid velocity threshold of $-50$ km s$^{-1}$ and/or
the detection of blue wings with $v_{84}$ $< -$300 km s$^{-1}$. The
{\em Herschel} data are not always sensitive to small-scale winds. Indeed,
small-scale molecular winds are known to exist in some of our
sources. For instance, \citet{Garcia2014} detect AGN-driven CO(3-2),
CO(6-5), HCN(4-3), HCO$^+$(4-3), and CS(7-6) outflows in NGC 1068 on
spatial scales of $\sim$~$20-35$~pc ($\sim0.3''-0.5''$), while OH is
observed purely in emission in the \emph{Herschel} data and is
therefore ambiguous regarding the existence of a large-scale outflow
until a full analysis of the velocity field is carried
out. Additionally, molecular outflows (or inflows) may be present, but
not centered on the central spaxel. For example, \emph{Herschel}
observations of OH 79, 84, and 65 $\mu$m, as well as HCN suggest the
existence of a spatially resolved outflow outside of the central
spaxel in NGC 3079 even though the central spaxel apparently displays
an OH inflow.  A detailed discussion of these objects is outside the
scope of this work, but these issues will be addressed in a future
paper.

\subsubsection{Driving Mechanisms of Molecular Outflows}

Correlations between the observed OH outflow velocities and the stellar masses, star formation rates, and AGN luminosities of the host galaxies can shed light upon the physical mechanisms responsible for driving molecular winds. There are good reasons to believe that energy and momentum injection from star formation activity plays a role in driving massive, galactic-scale outflows.  Previous studies  \citep[e.g.][]{Schwartz2004, RupkeSecond2005, Martin2005, Weiner2009, Cicone2014} have shown the existence of a positive trend between SFRs and outflow velocities. Curiously, V13 found no such trend in their ULIRGs + QSO sample. It appears, as V13 surmised, that the lack of a trend between these quantities was due to the limited range in SFR of the V13 sample, which spans only $\sim2$ dex. As seen in \autoref{fig:sfrVelocityV2}, when the BAT AGN are combined with the ULIRGs of V13, the sample spans a range $\sim3$ dex in SFR, and a positive correlation between outflow velocities and SFRs becomes apparent. 

Positive correlations between stellar masses and outflow velocities in the neutral and ionized phases of the ISM have also been reported in the literature \citep[e.g][]{RupkeSecond2005, Martin2005, Weiner2009}. V13 did not see this trend in the molecular data of their ULIRG + QSO sample, but again attributed this negative result to the small range in stellar masses of their sample ($\sim$ 1 dex). Our BAT AGN + ULIRG + QSO sample now extends over a stellar mass range of $\sim$ 2 dex and reveals a significant trend (see \autoref{fig:stellarMassVelocity}). However, note that this trend may be of secondary origin since a positive correlation is well known to exist between stellar mass and SFR in galaxies, both locally \citep[e.g. $0.015 \leq z \leq 0.1$;][]{Elbaz2007, Shimizu2015, Renzini2015} and at high redshift \citep[e.g. $0.7 < z < 1.5$;][]{Rubin2010, Whitaker2014} $-$ the so-called Main Sequence of star-forming galaxies. 

V13 reported a weak trend between wind velocities ($v_{50}$ and $v_{84}$) and AGN fractions in their ULIRG sample. They cautioned that the correlation could be merely an obscuration effect where both the AGN and central high-velocity outflowing material are more easily detectable when the dusty material has been swept away or is seen more nearly face-on. Indeed, by adding our BAT AGN sample to the ULIRG sample, we find that this weak correlation disappears. AGN fractions are thus not a good predictor of molecular outflows. On the other hand, a convincing causal connection was presented in V13 between AGN luminosities and molecular outflow velocities with a possible steepening of the relation above $\log(L^{break}_{\mathrm{AGN}}/L_\odot)=11.8\pm0.3$. Our results do indeed strongly suggest that at higher AGN luminosities ($\log(L_{\mathrm{AGN}}/L_\odot) \gtrsim 11.5$) the AGN dominates over star formation in driving the outflow. At lower luminosities ($\log(L_{\mathrm{AGN}}/L_\odot) \lesssim 11.5$) the AGN may not have the energetics required to drive fast molecular winds. 

Statistically, the correlation between wind velocity and AGN luminosity is stronger than the correlation between wind velocity and SFR (see \autoref{tab:statResults}). However, while the AGN may be the dominant source for driving the winds observed in our sample, it is clear that SFR also plays a role in driving these winds. The presence of an AGN seems to boost the observed velocity over that which would be observed in purely star forming systems \citep{Cicone2014}, although large scatter is observed (see \autoref{fig:velLumPlotWithUlirg}). This scatter may have multiple origins: (1) If the outflow is not spherically symmetric, projection effects will produce scatter in the observed velocities. (2) The efficiency to entrain material in the outflow depends on several complex factors associated with the acceleration mechanisms and the multi-phase nature of these processes. For instance, if radiation pressure is the dominant mechanism driving the outflow, radiative transfer effects are probably important so that not all of the OH-detected gas is ``seeing'' the same radiation field and therefore not experiencing the same radiation force. Similarly, if the dominant driving mechanism is ram pressure of a fast diffuse medium on dense cloudlets,  one might expect to observe a range of velocities depending on the characterisics (e.g., sizes and densities) of the cloudlets entrained in the flow and their survival timescales.  (3) The AGN luminosity is measured from the 14 -- 195 keV flux and therefore represent the current value of the AGN luminosity. In contrast, the observed outflow was likely produced $\sim 1$ to several $\times 10^6$ years ago based on the measured OH velocities and inferred outflow sizes ($\le 1$ kpc; \citet{Sturm2011, Gonzalez2014, Gonzalez2015}). AGN variability may therefore cause considerable scatter in \autoref{fig:velLumPlotWithUlirg}. 

\subsection{Inflows}

Seven objects (Centaurus A, Circinus, NGC 1125, NGC 3079, NGC 3281, NGC 4945,
and NGC 7582) have OH absorption features with median velocities
larger than 50 km s$^{-1}$, corresponding to an inflow detection rate
of $\sim$40\%. By far the best case for inflow in our sample is
Circinus where OH 119 $\mu$m shows an inverted P-Cygni
profile. Inverted P-Cygni profiles are also tentatively detected in
Centaurus~A and NGC 3281.

Interestingly, previous searches for neutral-gas (\ion{Na}{1}~D) outflows/inflows in IR-faint Seyfert galaxies have also shown distinctly higher detection rates of inflows than outflows. Specifically, in an analysis of 35 IR-faint Seyferts ($10^{9.9} < L_{\mathrm{IR}}/L_\odot < 10^{11.2}$), \citet{Krug2010} reported an inflow detection rate of $\sim 37\%$ and an outflow detection rate of only $\sim 11\%$. These numbers are similar to those derived here, and considerably different from those measured in (U)LIRGs using OH ($\sim 10\%$; V13) and \ion{Na}{1}~D \citep[e.g. $\sim 15\%$ inflow detection rate for 78 starbursting galaxies with $\log( L_\mathrm{IR}/L_\odot)=10.2-12.0$;][]{RupkeFirst2005, RupkeSecond2005}. The origin of this difference is unknown. The fast winds in (U)LIRGs may disturb the neutral-molecular gas and prevent it from infalling to the center. On average, IR-bright sources are also richer in gas and dust than IR-faint galaxies. This material may be masking the central regions where inflow is taking place. 

\subsection{The 9.7 $\mu$m Silicate Feature}

The analysis of the strength of the \silicate feature can provide insight into the mechanism responsible for the excitation of OH 119 $\mu$m observed in our sample. As seen in \autoref{fig:ohS97V2}, the comparison of the OH 119 $\mu$m equivalent width, EW$_\mathrm{OH}$, and the strength of the silicate feature, \silicate, implies a rather tight connection between OH gas and mid-infrared obscuration (a correlation is also found between OH 65 $\mu$m and \silicate; see also \citet{Gonzalez2015}). The clear trend of deeper OH 119 $\mu$m absorption and fainter OH 119 $\mu$m emission with increasing silicate obscuration (more positive values of \silicate) suggests that OH 119 $\mu$m emission is strongly affected by the obscuring geometry. Our results expand on those of V13 and S13 who found a similar trend with OH equivalent width and the strength of the 9.7 $\mu$m feature among ULIRGs. 

\begin{figure}
\centering
\label{fig:s97OhGalType2ObjDistV2}
\includegraphics[scale=0.4]{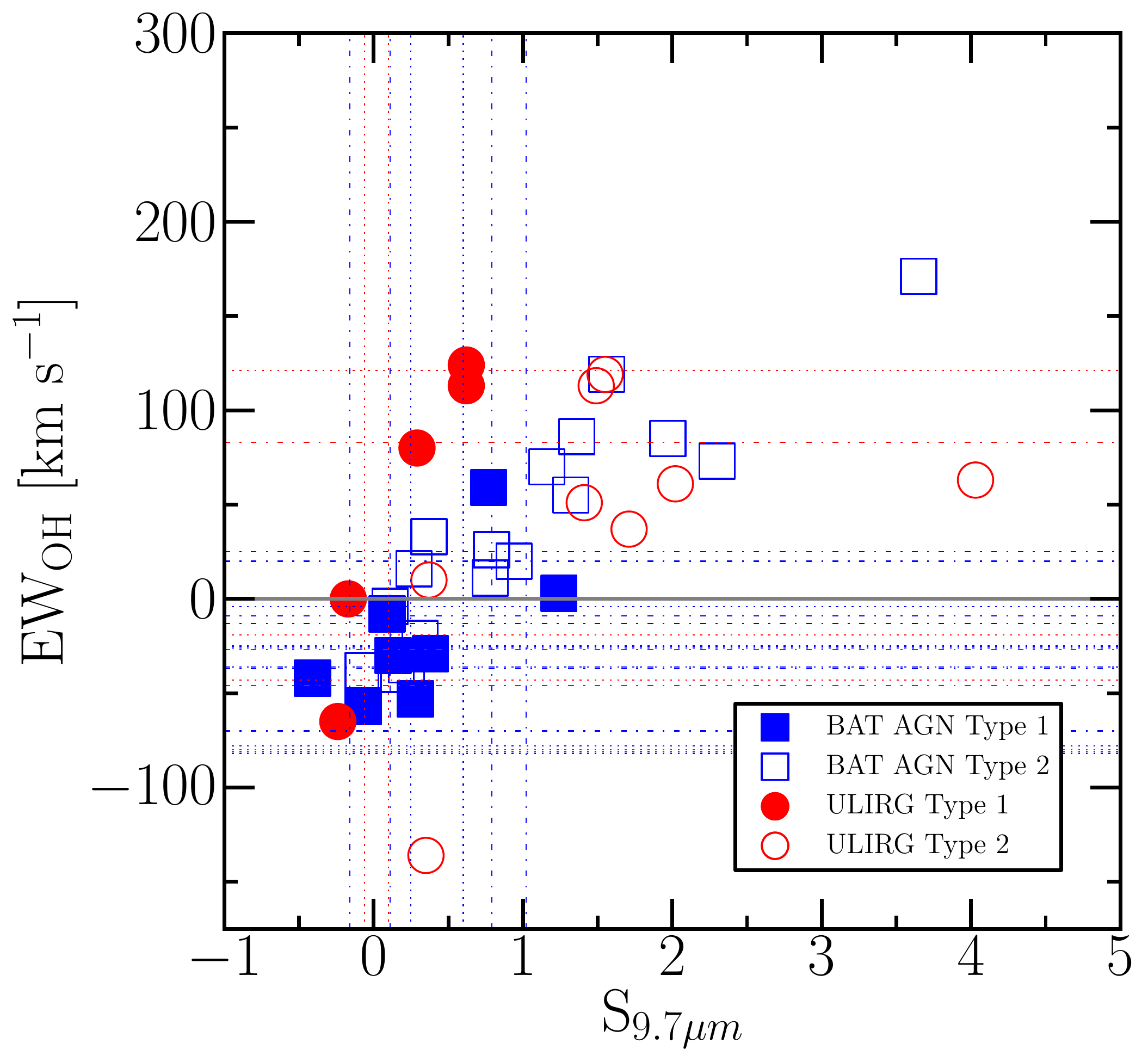}
\caption{Total (absorption + emission) equivalent widths of OH 119 $\mu$m as a function of the apparent strength of the 9.7 $\mu$m silicate feature relative to the local mid-infrared continuum. Note that \silicate is a logarithmic quantity and can be interpreted as the apparent silicate optical depth. The strength of this absorption feature increases to the right. Also note that objects classified as LINERs have been excluded from this plot. Filled markers refer to Type 1 and open markers refer to Type 2. Blue squares and red circles represent BAT AGN and ULIRGs/PG QSOs, respectively. Vertical lines represent objects in which OH was not detected. Horizontal lines represent objects with a null \silicate detection. Dotted lines and dash-dotted lines refer to Type 1 and Type 2, respectively. Blue and red lines indicate BAT AGN and ULIRGs/PG QSOs, respectively.}
\end{figure}

S13 argue that the OH 119 $\mu$m emission region often lies within the buried nucleus and that radiative excitation is the dominant source of OH 119 $\mu$m emission. In reality, the geometry of the silicate obscuration and the source of the OH 119 $\mu$m emission may be more complex. \autoref{fig:s97OhGalType2ObjDistV2} plots the total equivalent width of OH 119 $\mu$m as a function of \silicate and distinguishes between AGN spectral type for each object. Note that objects classified as LINERs have been excluded from this plot due to the ambiguous energy source in these objects \citep{Sturm2006}. None of the Type 1 galaxies (BAT AGN or ULIRGs) show a strong silicate absorption (\silicate $\gtrsim 1.5$). Interestingly, we see that OH 119 $\mu$m for some Type 2 BAT AGN and ULIRGs is observed in emission (EW$_{\mathrm{OH}} < 0 \text{ km s}^{-1}$) while the silicate feature, \silicate, is seen in absorption. It is possible that the OH 119 $\mu$m emission region is not nuclear, but is instead distributed throughout a circumnuclear starburst where the number density ($n$(H$_2) \sim$ a few times 10$^5$ cm$^{-3}$), temperature ($\sim100$ K), and OH abundance ($X$(OH) $\sim2\times10^{-6}$) are sufficient for collisional rather than radiative excitation  of the upper level of OH 119 $\mu$m \citep[e.g. NGC 1068; ][]{Spinoglio2005}. Until we examine in detail all 5 $\times$ 5 spaxels of the PACS data, the location of the OH emission in our objects will remain unclear. This type of analysis can be done in just a few select objects; this will be discussed in a future paper.

\begin{figure}
\centering
\label{fig:axRatioS97GalTypeV2}
\includegraphics[scale=0.4]{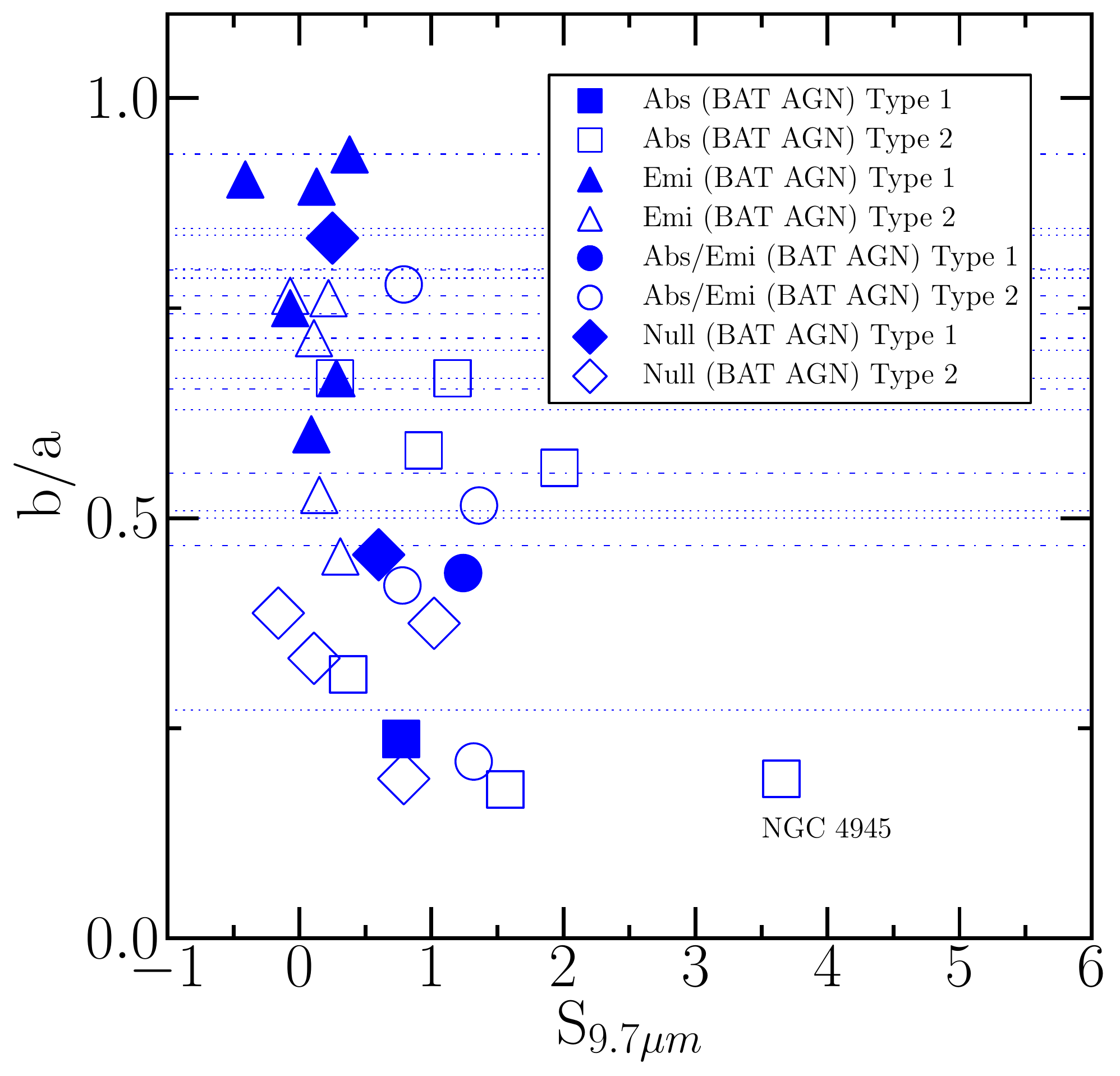}
\caption{Ratio of the semi-minor axis to the semi-major axis (a proxy for the inclination of the host galaxy disk) as a function of \silicate for the BAT AGN sample. Squares, triangles, and circles represent BAT AGN in which OH is observed purely in absorption, purely in emission, composite absorption/emission, respectively. Diamonds represent objects in which OH was undetected. Filled points and dash-dotted lines indicate Type 1 while open points and dotted lines indicate Type 2 AGN. Horizontal lines represent objects with a null 9.7 $\mu$m silicate feature detection.}
\end{figure}

Determining the origin of the silicate feature is also a challenge. The dust responsible for this feature may reside in the AGN torus, or on larger scale in the disk of the host galaxy, or some combination of these two. 
\autoref{fig:axRatioS97GalTypeV2} plots the ratio of the semi-minor and semi-major axes (a proxy for the inclination of the host galaxy disk) as a function of \silicate for BAT AGN. We have excluded the ULIRG/PG QSO sample from this particular analysis because these objects are undergoing or have undergone a major merger, and therefore, do not have a well-defined galactic plane or inclination. Visual inspection of \autoref{fig:axRatioS97GalTypeV2} suggests a weak trend between the inclination of the BAT AGN host galaxy and the depth of the 9.7 $\mu$m silicate feature. First, we see that OH for nearly face-on hosts ($b/a \gtrsim 0.8$) is seen only in pure emission and the strength of \silicate is weak. Second, we find that nearly edge-on galaxies show the broadest range of silicate absorption strengths, including the most extreme case of NGC 4945. However, the lack of a strong trend between inclination and \silicate strength suggests that the dust responsible for the 9.7 $\mu$m silicate feature is located not only in the plane of the host galaxy, but also in the nuclear torus. Our results are qualitatively consistent with \citet{Goulding2012}, who invoke a clumpy torus paradigm of many individual optically thick clouds \citep{Nenkova2002, Nenkova2008} and suggest that deeper silicate features (\silicate $\gtrsim 0.5$) are due to dust distributed at radii much larger ($\gg$ pc) than that predicted for a torus. \\

\section{Conclusions}
We present the results of our analysis of \emph{Herschel}/PACS spectroscopic observations of 52 nearby ($d<50$ Mpc) BAT AGN selected from the very hard X-ray (14-195 keV) \emph{Swift}-BAT Survey of local AGN. We also include in our analysis the \emph{Herschel}/PACS data on 38 ULIRGs and 5 PG QSOs from V13. The depth of the silicate feature at 9.7 $\mu$m feature in all these objects is measured from archival Spitzer/IRS data. Our combined BAT AGN + ULIRG + QSO sample covers a range of AGN luminosity and SFR of $\gtrsim3$ dex and $\sim3$ dex, respectively. The main results from our analysis are:

\begin{enumerate}
\item{The OH 119 $\mu$m feature is detected in 42 of the 52 objects in our BAT AGN sample. Of these detections, OH 119 $\mu$m is observed purely in emission for 25 targets, purely in absorption for 12 targets, and absorption+emission composite for 5 targets.}
\item{Evidence for molecular outflows (absorption profiles with median velocities more blueshifted than $-$50 km s$^{-1}$ and/or blueshifted wings with 84-percentile velocities less than $-$300 km s$^{-1}$) is seen in only four objects (NGC 7479, NGC 5506, NGC 7172, and IC 5063), corresponding to 24\% of all targets where this search for outflows was possible. This outflow detection rate is significantly smaller than that found in ULIRGs by V13 ($\sim$70\%). The best case for outflow is NGC 7479, an object with one of the highest infrared surface densities among our BAT AGN. }
\item{We find evidence for molecular inflows (absorption profiles with median velocities more redshifted than 50 km s$^{-1}$) in seven objects (Circinus, NGC 1125, NGC 3079, NGC 3281, and NGC 4945), corresponding to an inflow detection rate of $\sim$40\%, considerably higher than the rate measured among ULIRGs ($\sim$10\%) but similar to the detection rate of neutral-gas (\ion{Na}{1}~D) inflows among IR-faint Seyfert galaxies. By far the best case for OH inflow among our BAT AGN is Circinus, where OH 119 $\mu$m shows a distinct inverted P-Cygni profile.}
\item{The positive correlation between OH velocities and AGN luminosities reported in V13 is strengthened in the combined sample, but it is not seen in the BAT AGN sample. This suggests that luminous AGN play a dominant role in driving the fastest winds, but stellar processes will dominate when the AGN is weak or absent. }
\item{We confirm that the absorption strength of OH 119 $\mu$m is a good proxy for dust optical depth in these systems. Our findings are consistent with most, but not all, of the OH 119 $\mu$m emission originating near the AGN. Spatially resolved OH emission is seen in a few objects in our sample (e.g., NGC~1068) and may originate from a circumnuclear starburst where the number density and temperature are sufficiently high to collisionally excite the upper level of OH 119 $\mu$m.}
\item{A comparison of the strength of the 9.7 $\mu$m silicate feature with the inclination of the host galaxy disk and spectral type of the AGN confirm earlier results that the dust responsible for this feature is located not only in the nuclear component, but also in the disk of the host galaxy.}
\end{enumerate}

\acknowledgements Support for this work was provided by NASA through {\em Herschel} contracts 1427277 and 1454738 (M.S., S.V., and M.M.). We thank Lisa Winter, Michael Koss, Richard Mushotsky, Ranjan Vasudevan, Steve Hailey-Dunsheath, Ric Davies, Linda Tacconi, David Rupke, Jack Tueller, and Wayne Baumgartner who were Co-Investigators of the original OT-2 \emph{Herschel} proposal. We also thank Alessandra Contursi for her technical assistance and the referee whose thoughtful suggestions helped to improve this paper. This research made use of \emph{PySpecKit}, an open-source spectroscopic toolkit hosted at http://pyspeckit.bitbucket.org.  This work has made use of NASA's Astrophysics Data System Abstract Service and the NASA/IPAC Extragalactic Database (NED), which is operated by the Jet Propulsion Laboratory, California Institute of Technology, under contract with the National Aeronautics and Space Administration.

\bibliographystyle{apj}
\bibliography{ms}
\mbox{~}

\clearpage
\newpage
\appendix \label{app:appendix}
\counterwithin{figure}{section}
\section{\emph{Spitzer} MIR SPECTRA}    
\begin{figure*}[b!]
 \vspace{-0.3in} 
\centering
\label{fig:s97Profiles1}
\includegraphics[scale=0.92]{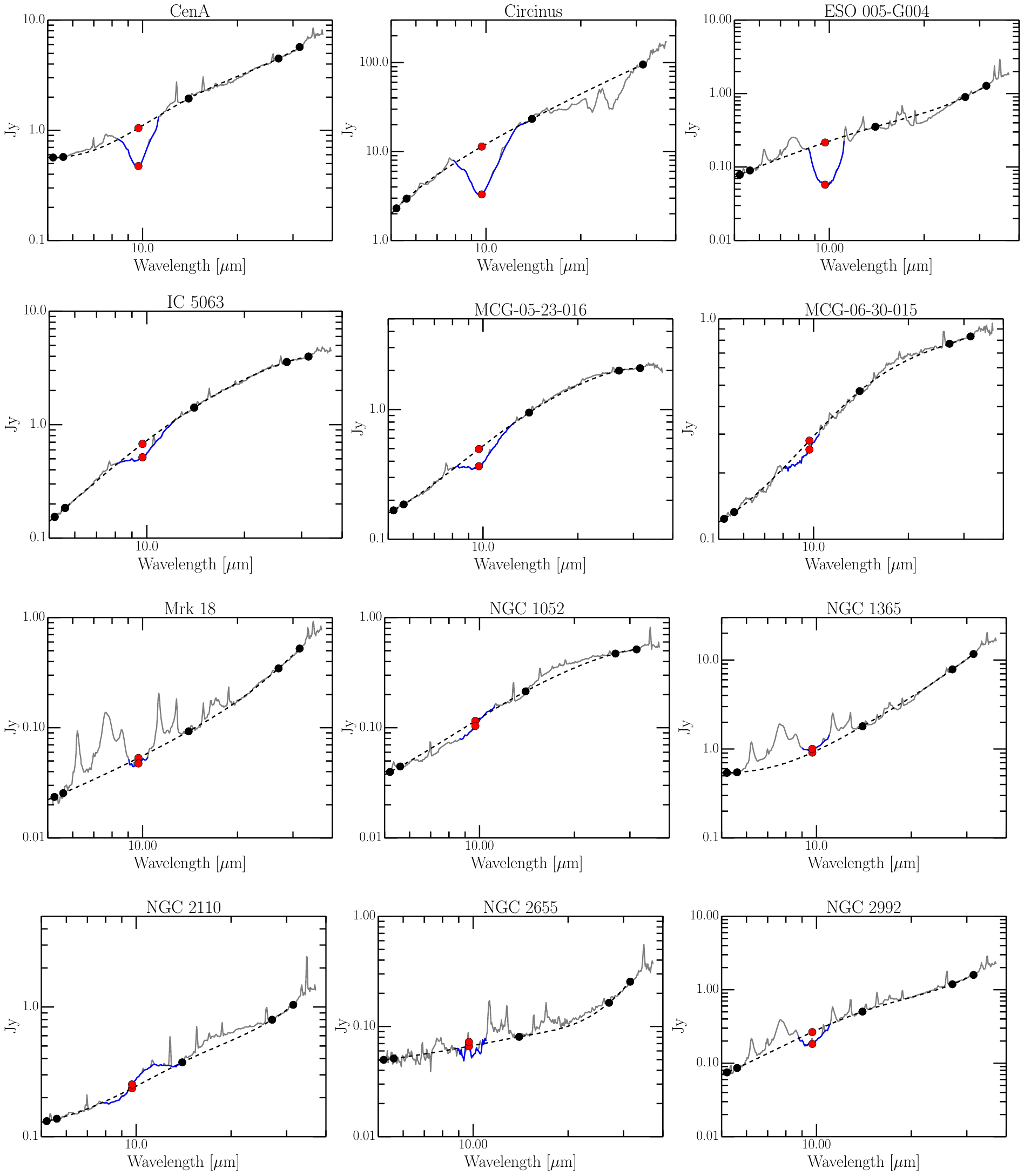}
\caption{Mid-infrared (5-37$\mu$m) spectra used to measure \silicate. The dashed line is the continuum calculated from the cubic spline interpolation fitted to the pivot points shown as black dots. Red dots show $f_{cont}(9.7\,\mu$m) (located on dashed continuum line) and  $f_{obs}(9.7\,\mu$m) (located on the solid black line or the observed flux density). The blue line shows the integration range used to calculate the flux and total equivalent width of the 9.7 $\mu$m silicate feature (see \autoref{tab:batSilicateProperties} and \autoref{tab:ulirgSilicateProperties}). \\
\\
(Figures of all MIR spectra are available in the online journal. A portion is shown here for guidance regarding form and content).}
\end{figure*}
\end{document}

%% file: tab1.tex
\capstartfalse
\begin{deluxetable}{r c c c}
\tablecolumns{4}
\scriptsize
\tabletypesize{\scriptsize}
\tablewidth{0pt}
\tablecaption{\emph{Herschel} Observations of BAT AGN
\label{tab:batAgnObservations}}
\tablehead{
\colhead{Name}	& \colhead{OBSID}	& 	\colhead{t$_{exp}$ [sec]}	& \colhead{Program}\\
\cline{1-4}\\
\colhead{(1)}    & \colhead{(2)} & \colhead{(3)}   & \colhead{(4)} 
}
\startdata
CenA	&	1342225989	&	976	&	OT1\_shaileyd\_1		\\
Circinus	&	1342225147	&	958	&	OT1\_shaileyd\_1		\\
ESO 005$-$G004 	&	1342245457	&	746	&	OT2\_sveilleu\_6		\\
ESO 137$-$34 	&	1342252089	&	1452	&	OT2\_sveilleu\_6		\\
IC 5063 	&	1342241848	&	2897	&	OT2\_sveilleu\_6		\\
IRAS 04410+2807 	&	1342249997	&	2897	&	OT2\_sveilleu\_6		\\
IRAS 19348$-$0619 	&	1342241495	&	1452	&	OT2\_sveilleu\_6		\\
MCG$-$05.23.16 	&	1342245454	&	14251	&	OT2\_sveilleu\_6		\\
MCG$-$06.30.15 	&	1342247813	&	14251	&	OT2\_sveilleu\_6		\\
Mrk18 	&	1342253541	&	4309	&	OT2\_sveilleu\_6		\\
NGC 1052 	&	1342247734	&	14251	&	OT2\_sveilleu\_6		\\
NGC 1068 	&	1342191154	&	3944	&	SHINING target		\\
NGC 1125 	&	1342247722	&	4309	&	OT2\_sveilleu\_6		\\
NGC 1365 	&	1342247546	&	746	&	OT2\_sveilleu\_6		\\
NGC 1566 	&	1342244440	&	746	&	OT2\_sveilleu\_6		\\
NGC 2110 	&	1342250314	&	2158	&	OT2\_sveilleu\_6		\\
NGC 2655 	&	1342246552	&	1452	&	OT2\_sveilleu\_6		\\
NGC 2992 	&	1342246246	&	746	&	OT2\_sveilleu\_6		\\
NGC 3079 	&	1342221391	&	8045	&	DDT\_esturm\_4		\\
NGC 3081 	&	1342245955	&	2897	&	OT2\_sveilleu\_6		\\
NGC 3227 	&	1342197796	&	2923	&	GT1\_lspinogl\_4		\\
NGC 3281 	&	1342248310	&	746	&	OT2\_sveilleu\_6		\\
NGC 3516 	&	1342245980	&	3128	&	GT1\_lspinogl\_6		\\
NGC 3718 	&	1342253721	&	7129	&	OT2\_sveilleu\_6		\\
NGC 3783 	&	1342247816	&	2897	&	OT2\_sveilleu\_6		\\
NGC 4051 	&	1342247512	&	1452	&	OT2\_sveilleu\_6		\\
NGC 4102 	&	1342247002	&	746	&	OT2\_sveilleu\_6		\\
NGC 4138 	&	1342256950	&	1452	&	OT2\_sveilleu\_6		\\
NGC 4151 	&	1342247511	&	746	&	OT2\_sveilleu\_6		\\
NGC 4258 	&	1342257242	&	746	&	OT2\_sveilleu\_6		\\
NGC 4388 	&	1342197911	&	3453	&	GT1\_lspinogl\_4		\\
NGC 4395 	&	1342247533	&	1452	&	OT2\_sveilleu\_6		\\
NGC 4579 	&	1342248535	&	746	&	OT2\_sveilleu\_6		\\
NGC 4593 	&	1342248372	&	1452	&	OT2\_sveilleu\_6		\\
NGC 4939 	&	1342248509	&	1452	&	OT2\_sveilleu\_6		\\
NGC 4941 	&	1342248508	&	2862	&	OT2\_sveilleu\_6		\\
NGC 4945 	&	1342247792	&	958	&	OT1\_shaileyd\_1		\\
NGC 5033 	&	1342247011	&	746	&	OT2\_sveilleu\_6		\\
NGC 5273 	&	1342246800	&	14251	&	OT2\_sveilleu\_6		\\
NGC 5290 	&	1342247535	&	2158	&	OT2\_sveilleu\_6		\\
NGC 5506 	&	1342247811	&	746	&	OT2\_sveilleu\_6		\\
NGC 5728 	&	1342249309	&	2508	&	GT1\_lspinogl\_6		\\
NGC 5899 	&	1342247007	&	746	&	OT2\_sveilleu\_6		\\
NGC 6221 	&	1342252088	&	746	&	OT2\_sveilleu\_6		\\
NGC 6300 	&	1342253357	&	746	&	OT2\_sveilleu\_6		\\
NGC 6814 	&	1342241849	&	746	&	OT2\_sveilleu\_6		\\
NGC 7172 	&	1342218490	&	2933	&	GT1\_lspinogl\_4		\\
NGC 7213 	&	1342245961	&	1452	&	OT2\_sveilleu\_6		\\
NGC 7314 	&	1342245225	&	746	&	OT2\_sveilleu\_6		\\
NGC 7465 	&	1342245963	&	1452	&	OT2\_sveilleu\_6		\\
NGC 7479 	&	1342258846	&	746	&	OT2\_sveilleu\_6		\\
NGC 7582 	&	1342257273	&	746	&	OT2\_sveilleu\_6		
\enddata
\tablecomments{Column 1: Galaxy name. Column 2: Observation ID. Column 3: Exposure time. Column 4: Program.}
\end{deluxetable}
\capstarttrue

%% file: tab2.tex
\capstartfalse 
\begin{deluxetable*}{r l c c c c c c c c}
\tablecolumns{10}
\tabletypesize{\scriptsize}
\tablewidth{0pt}
\tablecaption{Galaxy Properties
\label{tab:batAgnProperties}}
\tablehead{
\colhead{(Name)}    & \colhead{z} & \colhead{Distance} & \colhead{$\alpha_{\rm AGN}$}   & \colhead{log $L_{\rm AGN}$} & \colhead{log M$_{*}$} & \colhead{log SFR} & \colhead{$f_{cen}/f_{tot}$}&  \colhead{$f_{30\,\mu m}/f_{15\,\mu m}$}  & \colhead{Type} \\
\colhead{}    & \colhead{} & \colhead{(Mpc)} & \colhead{(\%)} & \colhead{($L_\odot$)}   & \colhead{(M$_\odot$)} & \colhead{(M$_\odot$ yr$^{-1}$)} & \colhead{} & \colhead{} & \colhead{} \\
\cline{1-10}\\
\colhead{(1)}    & \colhead{(2)} & \colhead{(3)}   & \colhead{(4)}    & \colhead{(5)} & \colhead{(6)} & \colhead{(7)} & \colhead{(8)} & \colhead{(9)} & \colhead{(10)}
}
\startdata
CenA	&	0.001901\tablenotemark{a}	&	3.7	&	93.2	&	9.78	&	\nodata	&	\nodata	&	0.14	&	2.44	&	Type 2	\\
Circinus	&	0.001573\tablenotemark{a}	&	4.2	&	86.6	&	9.19	&	\nodata	&	\nodata	&	0.22	&	3.32	&	Type 1	\\
ESO 005$-$G004 	&	0.006379\tablenotemark{b}	&	22.4	&	89.4	&	9.73	&	\nodata	&	0.72	&	0.15	&	2.95	&	Type 2	\\
ESO 137$-$34 	&	0.009144	&	33	&	\nodata	&	9.99	&	\nodata	&	0.5	&	0.2	&	\nodata	&	Type 2	\\
IC 5063 	&	0.011198\tablenotemark{c}	&	49	&	93.3	&	10.75	&	\nodata	&	0.61	&	0.38	&	2.43	&	Type 2	\\
IRAS 04410+2807 	&	0.011268	&	48.7	&	\nodata	&	10.61	&	\nodata	&	0.43	&	0.37	&	\nodata	&	Type 2	\\
IRAS 19348$-$0619 	&	0.010254	&	44.3	&	\nodata	&	10.17	&	\nodata	&	0.66	&	0.29	&	\nodata	&	Type 1	\\
MCG$-$05.23.16 	&	0.008486	&	36.6	&	96.9	&	10.94	&	9.56	&	$-$0.53	&	0.59	&	1.97	&	Type 2	\\
MCG$-$06.30.15 	&	0.007749	&	33.4	&	99.9	&	10.36	&	\nodata	&	$-$0.4	&	0.57	&	1.60	&	Type 1	\\
Mrk18 	&	0.011345\tablenotemark{d}	&	47.9	&	79.0	&	9.96	&	9.57	&	0.41	&	0.48	&	4.40	&	Type 2	\\
NGC 1052 	&	0.005037	&	19.5	&	95.7	&	9.56	&	10.35	&	\nodata	&	0.46	&	2.12	&	Type 2	\\
NGC 1068 	&	0.003931\tablenotemark{a}	&	12.7	&	100	&	9.26	&	\nodata	&	\nodata	&	0.14	&	1.20	&	Type 2	\\
NGC 1125 	&	0.010931	&	47.2	&	60.8	&	10.1	&	\nodata	&	0.39	&	0.42	&	7.31	&	Type 2	\\
NGC 1365 	&	0.005349\tablenotemark{a}	&	17.9	&	75.3	&	9.82	&	\nodata	&	1.51	&	0.18	&	4.96	&	Type 1	\\
NGC 1566 	&	0.005017	&	12.2	&	\nodata	&	9.01	&	\nodata	&	\nodata	&	0.21	&	\nodata	&	Type 1	\\
NGC 2110 	&	0.007789	&	35.6	&	94.0	&	11.11	&	10.63	&	0.49	&	0.43	&	2.34	&	Type 2	\\
NGC 2655 	&	0.00467	&	24.4	&	91.8	&	9.41	&	\nodata	&	$-$0.1	&	0.25	&	2.63	&	Type 2	\\
NGC 2992 	&	0.00771	&	31.6	&	92.1	&	9.94	&	10.31	&	0.75	&	0.38	&	2.59	&	Type 2	\\
NGC 3079 	&	0.003797\tablenotemark{d}	&	19.1	&	68.7	&	9.59	&	9.98	&	1.22	&	0.29	&	5.99	&	Type 2	\\
NGC 3081 	&	0.007976	&	26.5	&	93.8	&	10.27	&	10.31	&	0.12	&	0.35	&	2.37	&	Type 2	\\
NGC 3227 	&	0.004001\tablenotemark{a}	&	18.7	&	88.6	&	10.09	&	9.98	&	0.55	&	0.33	&	3.05	&	Type 1	\\
NGC 3281 	&	0.010674	&	46.1	&	92.7	&	10.77	&	10.24	&	0.84	&	0.46	&	2.51	&	Type 2	\\
NGC 3516 	&	0.008889\tablenotemark{e}	&	52.5	&	95.5	&	11.02	&	10.46	&	0.19	&	0.44	&	2.15	&	Type 1	\\
NGC 3718 	&	0.003312	&	17	&	\nodata	&	9.05	&	9.98	&	$-$0.3	&	0.23	&	\nodata	&	LINER	\\
NGC 3783 	&	0.009791\tablenotemark{f}	&	47.8	&	95.2	&	11.12	&	\nodata	&	0.68	&	0.11	&	2.19	&	Type 1	\\
NGC 4051 	&	0.002490\tablenotemark{a}	&	14.3	&	95.2	&	9.41	&	9.44	&	0.57	&	0.34	&	2.19	&	Type 1	\\
NGC 4102 	&	0.002823	&	20.4	&	61.7	&	9.57	&	9.68	&	1.02	&	0.51	&	7.15	&	LINER	\\
NGC 4138 	&	0.002962	&	20.7	&	99.3	&	9.62	&	9.61	&	0.04	&	0.05	&	1.67	&	Type 1	\\
NGC 4151 	&	0.003502\tablenotemark{a}	&	9.9	&	100	&	10.23	&	\nodata	&	$-$0.38	&	0.24	&	1.48	&	Type 1	\\
NGC 4258 	&	0.001494	&	7.5	&	98.4	&	8.59	&	\nodata	&	\nodata	&	0.07	&	1.79	&	Type 2	\\
NGC 4388 	&	0.008467\tablenotemark{a}	&	21	&	84.3	&	10.6	&	10.53	&	0.58	&	0.14	&	3.64	&	Type 2	\\
NGC 4395 	&	0.001064	&	4.5	&	72.7	&	8.21	&	8.28	&	\nodata	&	0.1	&	5.34	&	Type 1	\\
NGC 4579 	&	0.00506	&	19.6	&	\nodata	&	9.11	&	\nodata	&	\nodata	&	0.21	&	\nodata	&	Type 2	\\
NGC 4593 	&	0.008455\tablenotemark{a}	&	30.8	&	76.5	&	10.43	&	10.75	&	\nodata	&	0.4	&	4.78	&	Type 1	\\
NGC 4939 	&	0.010374	&	44.8	&	90.2	&	10.21	&	\nodata	&	0.82	&	0.06	&	2.84	&	Type 2	\\
NGC 4941 	&	0.003707\tablenotemark{d}	&	18.7	&	89.5	&	9.36	&	\nodata	&	$-$0.19	&	0.22	&	2.93	&	Type 2	\\
NGC 4945 	&	0.001836\tablenotemark{g}	&	4.1	&	10.4	&	9.18	&	\nodata	&	\nodata	&	0.41	&	18.96	&	Type 2	\\
NGC 5033 	&	0.003211\tablenotemark{a}	&	19.6	&	\nodata	&	8.84	&	\nodata	&	0.95	&	0.1	&	\nodata	&	Type 1	\\
NGC 5273 	&	0.003619	&	16	&	77.4	&	9.06	&	9.64	&	\nodata	&	0.55	&	4.64	&	Type 1	\\
NGC 5290 	&	0.008583	&	35	&	\nodata	&	9.88	&	10.23	&	0.62	&	0.12	&	\nodata	&	Type 2	\\
NGC 5506 	&	0.006228\tablenotemark{a}	&	23.8	&	93.3	&	10.64	&	10.02	&	0.25	&	0.52	&	2.43	&	Type 1	\\
NGC 5728 	&	0.009475\tablenotemark{a}	&	30.6	&	70.4	&	10.43	&	10.78	&	0.78	&	0.41	&	5.71	&	Type 2	\\
NGC 5899 	&	0.008546	&	38.1	&	91.0	&	9.97	&	10.28	&	0.95	&	0.13	&	2.73	&	Type 2	\\
NGC 6221 	&	0.004999	&	12.2	&	65.4	&	8.99	&	\nodata	&	0.76	&	0.24	&	6.53	&	Type 2	\\
NGC 6300 	&	0.003699	&	14.4	&	87.0	&	9.82	&	\nodata	&	0.63	&	0.31	&	3.27	&	Type 2	\\
NGC 6814 	&	0.005214	&	22.8	&	98.0	&	10.11	&	10.3	&	0.6	&	0.05	&	1.83	&	Type 1	\\
NGC 7172 	&	0.009180\tablenotemark{e}	&	33.9	&	92.4	&	10.8	&	\nodata	&	0.83	&	0.24	&	2.54	&	Type 2	\\
NGC 7213 	&	0.005983\tablenotemark{d}	&	22	&	100	&	9.82	&	\nodata	&	0.25	&	0.12	&	1.46	&	Type 1	\\
NGC 7314 	&	0.004839\tablenotemark{h}	&	18.6	&	89.0	&	9.77	&	10.06	&	\nodata	&	0.13	&	3.00	&	Type 1	\\
NGC 7465 	&	0.006666\tablenotemark{i}	&	27.2	&	\nodata	&	9.54	&	\nodata	&	0.21	&	0.35	&	\nodata	&	Type 2	\\
NGC 7479 	&	0.007942	&	33.9	&	83.7	&	9.88	&	\nodata	&	1.14	&	0.37	&	3.73	&	Type 2	\\
NGC 7582 	&	0.005541\tablenotemark{j}	&	20.9	&	73.8	&	10.06	&	\nodata	&	1.2	&	0.49	&	5.18	&	Type 2		
\enddata
\tablenotetext{a}{Redshift is calculated using the [OI] 63.18 $\mu$m , [OI] 145.53 $\mu$m, [CII] 157.74 $\mu$m, and [NII] 121.90 $\mu$m emission lines.}
\tablenotetext{b}{Redshift is calculated using the [OI] 145.53 $\mu$m emission line.}
\tablenotetext{c}{Redshift is calculated using the [OI] 63.18 $\mu$m and [CII] 157.74 $\mu$m emission lines.}
\tablenotetext{d}{Redshift is calculated using the [CII] 157.74 $\mu$m emission line.}
\tablenotetext{e}{Redshift is calculated using the [CII] 157.74 $\mu$m and [NII]  121.90 $\mu$m emission lines.}
\tablenotetext{f}{Redshift is calculated using the [OI] 63.18 $\mu$m and [OI] 145.53 $\mu$m emission lines.}
\tablenotetext{g}{Redshift is calculated using the [OI] 63.18 $\mu$m, [OI] 145.53 $\mu$m, and [CII] 157.74 $\mu$m emission lines.}
\tablenotetext{h}{Redshift is calculated using the [OI] 145.53 $\mu$m and [CII] 157.74 $\mu$m emission lines.}
\tablenotetext{i}{Redshift is calculated using the [OI] 63.18 $\mu$m, [CII] 157.74 $\mu$m, and [NII] 121.90 $\mu$m emission lines.}
\tablenotetext{j}{Redshift is calculated using the [OI] 14.53 $\mu$m , [CII] 157.74 $\mu$m, and [NII] 121.90 $\mu$memission lines.}
\tablecomments{Column 1: Galaxy name. Column 2: Redshift value (When available, redshift is calculated from emission lines. Otherwise, redshifts are from NED). Column 3: When available mean, redshift-independent distance is obtained from NED. Otherwise, the luminosity distance is calculated via \citet{Wright2006}. Column 4: $\alpha_{\mathrm{AGN}}$, fractional contribution of the AGN to the bolometric luminosity; see \autoref{sec:galaxyProperties}. Column 5: AGN luminosity. Column 6: stellar masses are adopted from \citet{Koss2011}. Column 7: star formation rate; see \autoref{sec:galaxyProperties}. Column 8: Continuum flux density ratio at 119 $\mu$m of the central spaxel to all 25 spaxels. For reference, the average continuum ratio for a point source calculated from the five PG QSOs in our ULIRG sample is $f_{cen}/f_{tot}=0.56$. Column 9: 30 $\mu$m to 15 $\mu$m continuum flux density ratio. Column 10: Spectral type.}
\end{deluxetable*}
\capstarttrue

%% file: tab3.tex
\capstartfalse
\begin{deluxetable*}{rccccccccc}[ht!]
\tabletypesize{\scriptsize}
\tablecolumns{10}
\tablewidth{0pt}
\tablecaption{Properties of the OH 119 $\mu$m Profiles
\label{tab:ohProperties}}
\tablehead{
\colhead{(Name)}    & \colhead{v$_{50}$ (abs)} & \colhead{v$_{84}$ (abs)}   & \colhead{Flux$_{\rm abs}$} & \colhead{EQW$_{\rm abs}$} &  \colhead{v$_{50}$ (emi)} &
\colhead{v$_{84}$ (emi)}    & \colhead{Flux$_{\rm emi}$}   &  \colhead{EQW$_{\rm emi}$} & \colhead{EQW$_{\mathrm{Total}}$}\\
\colhead{}    & \colhead{(km s$^{-1}$)} & \colhead{(km s$^{-1}$)} & \colhead{(Jy km s$^{-1}$)}   & \colhead{(km s$^{-1}$)}    & \colhead{(km s$^{-1}$)} &
\colhead{(km s$^{-1}$)}    & \colhead{(Jy km s$^{-1}$)} & \colhead{(km s$^{-1}$)}  & \colhead{(km s$^{-1}$)}\\
\cline{1-10}\\
\colhead{(1)}    & \colhead{(2)} & \colhead{(3)}   & \colhead{(4)}    & \colhead{(5)} &
\colhead{(6)}    & \colhead{(7)}   & \colhead{(8)}    & \colhead{(9)}    & \colhead{(10)} 
}
\startdata
CenA	&	63	&	195	&	521.1	&	27	&	$-$201:	&	$-$327:	&	$-$107.9	&	$-$6	&	21\\
Circinus	&	81	&	201	&	494.3	&	8	&	$-$225	&	$-$327:	&	$-$313.9:	&	$-$5:	&	3\\
ESO 005$-$G004	&	9	&	$-$141	&	167.1	&	70	&	622	&	754	&	$-$32	&	$-$14	&	55\\
ESO 137$-$34	&	\nodata	&	\nodata	&	\nodata	&	\nodata	&	\nodata	&	\nodata	&	$-$9.4	&	$-$14	&	$-$14\\
IC 5063	&	$-$33::	&	$-$309::	&	53.7:	&	16	&	\nodata	&	\nodata	&	\nodata	&	\nodata	&	16\\
IRAS 04410+2807	&	\nodata	&	\nodata	&	\nodata	&	\nodata	&	21::	&	135::	&	$-$36.8::	&	$-$37::	&	$-$37::\\
IRAS 19348$-$0619	&	\nodata	&	\nodata	&	\nodata	&	\nodata	&	57	&	207	&	$-$28.5	&	$-$25:	&	$-$25:\\
MCG$-$05$-$23$-$016	&	\nodata	&	\nodata	&	\nodata	&	\nodata	&	57::	&	207::	&	$-$8.6::	&	$-$21::	&	$-$21::\\
MCG$-$06$-$30$-$015	&	\nodata	&	\nodata	&	\nodata	&	\nodata	&	129::	&	255::	&	$-$6.9::	&	$-$8::	&	$-$8::\\
Mrk 18	&	\nodata	&	\nodata	&	12.4:	&	6:	&	\nodata	&	\nodata	&	\nodata	&	\nodata	&	6\\
NGC 1052	&	\nodata	&	\nodata	&	\nodata	&	\nodata	&	$-$39::	&	117::	&	$-$9.9::	&	$-$27::	&	$-$27::\\
NGC 1068	&	\nodata	&	\nodata	&	\nodata	&	\nodata	&	15	&	213	&	$-$2479.1	&	$-$70	&	$-$70\\
NGC 1125	&	123:	&	$-$3:	&	24.8:	&	20	&	\nodata	&	\nodata	&	\nodata	&	\nodata	&	20\\
NGC 1365	&	\nodata	&	\nodata	&	\nodata	&	\nodata	&	177:	&	267:	&	$-$104.8:	&	$-$4:	&	$-$4:\\
NGC 1566	&	\nodata	&	\nodata	&	\nodata	&	\nodata	&	51	&	177	&	$-$93.6	&	$-$36	&	$-$36\\
NGC 2110	&	\nodata	&	\nodata	&	\nodata	&	\nodata	&	75	&	351	&	$-$103.9	&	$-$38	&	$-$38\\
NGC 2655	&	\nodata	&	\nodata	&	\nodata	&	\nodata	&	9::	&	135::	&	$-$8::	&	$-$9::	&	$-$9::\\
NGC 2992	&	45	&	$-$129	&	146.9	&	33	&	\nodata	&	\nodata	&	\nodata	&	\nodata	&	33\\
NGC 3079	&	87	&	$-$93	&	2580.2	&	119	&	\nodata	&	\nodata	&	\nodata	&	\nodata	&	119\\
NGC 3081	&	\nodata	&	\nodata	&	\nodata	&	\nodata	&	93	&	219	&	$-$38.7	&	$-$33	&	$-$33\\
NGC 3227	&	\nodata	&	\nodata	&	\nodata	&	\nodata	&	$-$27	&	147	&	$-$225.3	&	$-$53	&	$-$53\\
NGC 3281	&	273	&	429	&	313.9	&	99	&	$-$321	&	$-$453	&	$-$32	&	$-$10	&	$-$86\\
NGC 3516	&	\nodata	&	\nodata	&	\nodata	&	\nodata	&	$-$99	&	39	&	$-$52.4	&	$-$78	&	$-$78\\
NGC 3718	&	\nodata	&	\nodata	&	\nodata	&	\nodata	&	\nodata	&	\nodata	&	$-$9.3:	&	$-$15:	&	$-$15:\\
NGC 3783	&	\nodata	&	\nodata	&	\nodata	&	\nodata	&	$-$69::	&	81::	&	$-$16.8::	&	$-$33::	&	$-$33::\\
NGC 4051	&	\nodata	&	\nodata	&	\nodata	&	\nodata	&	$-$9:	&	141:	&	$-$102.2:	&	$-$57:	&	$-$57:\\
NGC 4102	&	$-$3	&	$-$135	&	343.4	&	12:	&	\nodata	&	\nodata	&	\nodata	&	\nodata	&	12:\\
NGC 4138	&	\nodata	&	\nodata	&	\nodata	&	\nodata	&	$-$99::	&	33::	&	$-$19.2::	&	$-$81::	&	$-$81::\\
NGC 4151	&	\nodata	&	\nodata	&	\nodata	&	\nodata	&	75	&	333:	&	$-$96.7	&	$-$82	&	$-$82\\
NGC 4258	&	\nodata	&	\nodata	&	\nodata	&	\nodata	&	\nodata	&	\nodata	&	$-$19.2	&	$-$18	&	$-$18\\
NGC 4388	&	\nodata	&	\nodata	&	32.7	&	18	&	\nodata	&	\nodata	&	\nodata	&	\nodata	&	18\\
NGC 4395	&	\nodata	&	\nodata	&	\nodata	&	\nodata	&	\nodata	&	\nodata	&	$-$14.2::	&	$-$254::	&	$-$254::\\
NGC 4579	&	\nodata	&	\nodata	&	\nodata	&	\nodata	&	$-$15::	&	111::	&	$-$40.3::	&	$-$36::	&	$-$36::\\
NGC 4593	&	\nodata	&	\nodata	&	\nodata	&	\nodata	&	15	&	141	&	$-$44.4	&	$-$26:	&	$-$26:\\
NGC 4939	&	\nodata	&	\nodata	&	\nodata	&	\nodata	&	\nodata	&	\nodata	&	$-$11.1	&	$-$45	&	$-$45\\
NGC 4941	&	\nodata	&	\nodata	&	\nodata	&	\nodata	&	3::	&	129::	&	$-$22.4::	&	$-$40::	&	$-$40::\\
NGC 4945	&	87	&	$-$51	&	66839.9	&	171	&	\nodata	&	\nodata	&	\nodata	&	\nodata	&	171\\
NGC 5033	&	\nodata	&	\nodata	&	\nodata	&	\nodata	&	\nodata	&	\nodata	&	$-$15.1	&	$-$4	&	$-$4\\
NGC 5273	&	\nodata	&	\nodata	&	\nodata	&	\nodata	&	87:	&	231:	&	$-$13.6:	&	$-$30:	&	$-$30:\\
NGC 5290	&	9::	&	$-$117::	&	14::	&	25::	&	\nodata	&	\nodata	&	\nodata	&	\nodata	&	25::\\
NGC 5506	&	45::	&	$-$357::	&	234:	&	59:	&	\nodata	&	\nodata	&	\nodata	&	\nodata	&	59:\\
NGC 5728	&	$-$15::	&	$-$141::	&	96::	&	20::	&	\nodata	&	\nodata	&	\nodata	&	\nodata	&	20::\\
NGC 5899	&	\nodata	&	\nodata	&	\nodata	&	\nodata	&	\nodata	&	\nodata	&	$-$11:	&	$-$15:	&	$-$15:\\
NGC 6221	&	\nodata	&	\nodata	&	\nodata	&	\nodata	&	$-$99::	&	33::	&	$-$32::	&	$-$4::	&	$-$4::\\
NGC 6300	&	$-$3	&	$-$201	&	301.5	&	70	&	\nodata	&	\nodata	&	\nodata	&	\nodata	&	70\\
NGC 6814	&	\nodata	&	\nodata	&	\nodata	&	\nodata	&	105::	&	231::	&	$-$33.8::	&	$-$82::	&	$-$82::\\
NGC 7172	&	$-$51:	&	$-$207::	&	252.6	&	85	&	\nodata	&	\nodata	&	\nodata	&	\nodata	&	85\\
NGC 7213	&	\nodata	&	\nodata	&	\nodata	&	\nodata	&	$-$3::	&	141::	&	$-$28.3::	&	$-$42::	&	$-$42::\\
NGC 7314	&	\nodata	&	\nodata	&	18	&	24	&	\nodata	&	\nodata	&	\nodata	&	\nodata	&	24\\
NGC 7465	&	\nodata	&	\nodata	&	\nodata	&	\nodata	&	$-$69::	&	57::	&	$-$25.6::	&	$-$13::	&	$-$13::\\
NGC 7479	&	$-$51	&	$-$658::	&	379.6	&	73	&	\nodata	&	\nodata	&	\nodata	&	\nodata	&	73\\
NGC 7582	&	105	&	3	&	383.6	&	13	&	604:	&	700::	&	$-$50.1::	&	$-$2	&	11
\enddata
\tablecomments{Column 1: galaxy name. Column 2: $v_{50}$(abs) is the median velocity of the fitted absorption profile {\em i.e.} 50\% of the absorption takes place at velocities above - more positive than - $v_{50}$. Column 3: $v_{84}$(abs) is the velocity above which 84\%  of the absorption takes place. Column 4: the total integrated flux for the absorption component(s). Column 5: the total equivalent width for the absoprtion component(s). Column 6: $v_{50}$ (emi) is the median velocity of the fitted emission profile. Column 7: $v_{84}$ (emi) is the velocity below which 84\% of the emission takes place. Column 8: the total integrated flux for the emission component(s). Column 9: the total equivalent width for the emission component(s). Column 10: The total equivalent width for the sum of the two components for one line of the OH doublet. Fluxes followed by a colon indicate uncertainties between 20\% and 50\%. Velocities followed by a colon indicate uncertainties between $50-150$ km s$^{-1}$.  Fluxes followed by a double colon indicate uncertainties $> 50\%$. Velocities followed by a double colon indicate uncertainties $>150$ km s$^{-1}$.}
\end{deluxetable*}
\capstarttrue

%% file: tab4.tex
\capstartfalse
\begin{deluxetable*}{rccccc}[t!]
\tabletypesize{\scriptsize}
\tablecolumns{6}
\tablewidth{0pt}
\tablecaption{$S_{9.7\,\mu\mathrm{m}}\,$ Continuum Parameters and Measured Values of the BAT AGN Sample
\label{tab:batSilicateProperties}}
\tablehead{
\colhead{(Name)}    & \colhead{Pivot Points} & \colhead{Integration Range} & \colhead{Flux$_{S_{9.7\,\mu\mathrm{m}}}$} & \colhead{EQW$_{S_{9.7\,\mu\mathrm{m}}}$} &  \colhead{$S_{9.7\,\mu\mathrm{m}}\,$} \\
\colhead{}    & \colhead{($\mu$m)} &  \colhead{($\mu$m)} & \colhead{(Jy $\mu$m)} & \colhead{($\mu$m)}   & \colhead{}\\
\cline{1-6}\\
\colhead{(1)}    & \colhead{(2)} & \colhead{(3)}   & \colhead{(4)}    & \colhead{(5)} & \colhead{(6)}
}
\startdata
CenA	&	5.2, 5.6, 14.0, 27.0, 31.5	&	8.3,	11.3	&	1.06	&	1.01	&	0.79\\ 
Circinus	&	5.2, 5.6, 14.0, 31.5	&	7.8,	14	&	23.38	&	2.06	&	1.24\\ 
ESO 005$-$G004	&	5.2, 5.6, 14.0, 27.0, 31.5	&	8.6,	11.19	&	0.34	&	1.55	&	1.32\\ 
ESO 137$-$34	&	\nodata	&		\nodata	&	\nodata	&	\nodata	&	\nodata\\
IC 5063	&	5.2, 5.6, 14.0, 27.0, 31.5	&	8.15, 12.3	&	0.37	&	0.55	&	0.27\\ 
IRAS 00410+2807	&	\nodata	&		\nodata	&	\nodata	&	\nodata	&	\nodata\\ 
IRAS 19348$-$0619	&	\nodata	&		\nodata	&	\nodata	&	\nodata	&	\nodata\\ 
MGC$-$05.23.16	&	5.2, 5.6, 14.0, 27.0, 31.5	&	8,	12.45	&	0.32	&	0.64	&	0.31\\ 
MGC$-$06.30.15	&	5.2, 5.6, 14.0, 27.0, 31.5	&	8.09,	10.4	&	0.05	&	0.16	&	0.09\\ 
Mrk 18	&	5.2, 5.6, 14.0, 27.0, 31.5	&	9,	10.4	&	0.01	&	0.12	&	0.11\\ 
NGC 1052	&	\nodata	&		\nodata	&	\nodata	&	\nodata	&	\nodata\\
NGC 1068	&	\nodata	&		\nodata	&	\nodata	&	\nodata	&	\nodata\\ 
NGC 1125		&	\nodata	&		\nodata	&	\nodata	&	\nodata	&	\nodata\\
NGC 1365	&	\nodata	&		\nodata	&	\nodata	&	\nodata	&	\nodata\\
NGC 1566	&	\nodata	&		\nodata	&	\nodata	&	\nodata	&	\nodata\\ 
NGC 2110		&	5.2, 5.6, 14.0, 27.0, 31.5	&	7.8,	14	&	0.17	&	0.74	&	$-$0.07\\
NGC 2655	&	\nodata	&		\nodata	&	\nodata	&	\nodata	&	\nodata\\
NGC 2992	&	5.2, 5.6, 14.0, 27.0, 31.5	&	8.75,	11.1	&	0.13	&	0.48	&	0.37\\
NGC 3079	&	5.2, 5.6, 14.0, 27.0, 31.5	&	8.8,	11.13	&	0.66	&	1.5	&	1.56\\
NGC 3081	&	5.2, 5.6, 14.0, 27.0, 31.5	&	8,	11.3	&	0.09	&	0.44	&	0.22\\
NGC 3227	&	5.2, 5.6, 14.0, 27.0, 31.5	&	8.75,	11.09	&	0.15	&	0.46	&	0.28\\
NGC 3281	&	5.2, 5.6, 7.8, 14.0, 27.0, 31.5	&	7.8,	14	&	1.98	&	2.27	&	1.36\\
NGC 3516	&	\nodata	&		\nodata	&	\nodata	&	\nodata	&	\nodata\\ 
NGC 3718	&	\nodata	&			&	\nodata	&	\nodata	&	\nodata\\ 
NGC 3783	&	5.2, 5.6, 14.0, 27.0, 31.5	&	7.9,	11.3	&	0.17	&	0.33	&	0.13\\
NGC 4051	&	5.2, 5.6, 14.0, 21.9, 24.9	&	8,	10.8	&	0.07	&	0.24	&	$-$0.07\\
NGC 4102	&	5.2, 5.6, 14.0, 27.0, 31.5	&	9,	10.95	&	0.23	&	0.52	&	0.41\\
NGC 4138	&	\nodata	&		\nodata	&	\nodata	&	\nodata	&	\nodata\\ 
NGC 4151	&	\nodata	&	7.8,	14	&	\nodata	&	\nodata	&	\nodata\\
NGC 4258	&	5.2, 5.6, 14.0, 27.0, 31.5	&	7.8,	13.39	&	0.14	&	1.15	&	$-$0.16\\
NGC 4388	&	5.2, 5.6, 14.0, 27.0, 31.5	&	8.2,	12.2	&	0.43	&	1.35	&	0.79\\ 
NGC 4395	&	5.2, 5.6, 14.0, 27.0, 31.5	&	7.8,	11.09	&	0.004	&	0.67	&	0.25\\
NGC 4579	&	\nodata	&		\nodata	&	\nodata	&	\nodata	&	\nodata\\ 
NGC 4593	&	\nodata	&		\nodata	&	\nodata	&	\nodata	&	\nodata\\
NGC 4939	&	\nodata	&		\nodata	&	\nodata	&	\nodata	&	\nodata\\
NGC 4941	&	5.2, 5.6, 7.8, 14.0, 27.0, 31.5	&	8,	11.13	&	0.02	&	0.33	&	0.15\\
NGC 4945	&	7.8, 14.0, 27.0, 31.5	&	7.8,	11.09	&	1.95	&	2.72	&	3.65\\
NGC 5033	&	\nodata	&		\nodata	&	\nodata	&	\nodata	&	\nodata\\ 
NGC 5273	&	4.6, 14.0, 27.0, 31.5	&	7.7,	11.05	&	0.01	&	0.66	&	0.38\\
NGC 5290	&	\nodata	&		\nodata	&	\nodata	&	\nodata	&	\nodata\\ 
NGC 5506	&	5.2, 5.6, 14.0, 27.0, 31.5	&	8.35,	12.15	&	1.2	&	1.16	&	0.77\\
NGC 5728	&	5.2, 5.6, 14.0, 27.0, 31.5	&	8.63,	11.16	&	0.15	&	1.17	&	0.94\\
NGC 5899	&	5.2, 5.6, 14.0, 27.0, 31.5	&	8.1,	11.15	&	0.08	&	1.44	&	1.02\\
NGC 6221	&	5.2, 5.6, 14.0, 27.0, 31.5	&	9.2,	10.7	&	0.03	&	0.13	&	0.11\\
NGC 6300	&	5.2, 5.6, 14.0, 27.0, 31.5	&	8.58,	12.2	&	0.67	&	1.71	&	1.16\\
NGC 6814	&	\nodata	&		\nodata	&	\nodata	&	\nodata	&	\nodata\\ 
NGC 7172	&	5.2, 5.6, 14.0, 27.0, 31.5	&	8.15,	12.52	&	0.68	&	2.37	&	1.97\\
NGC 7213	&	5.2, 5.6, 14.0, 27.0, 31.5	&	8.76,	13.3	&	0.36	&	2.26	&	$-$0.41\\
NGC 7314	&	5.2, 5.6, 7.8, 14.0	&	7.8,	14	&	0.09	&	1.38	&	0.6\\
NGC 7465	&	\nodata	&		\nodata	&	\nodata	&	\nodata	&	\nodata\\ 
NGC 7479	&	4.6, 7.8, 14.0, 27.0, 31.5	&	7.8,	12.9	&	1.98	&	3.09	&	2.3\\
NGC 7582	&	5.2, 5.6, 14.0, 27.0, 31.5	&	8.8,	11.05	&	0.81	&	0.95	&	0.78\\
\enddata
\tablecomments{Column 1: galaxy name. Column 2: Pivot points in microns used for the continuum interpolation. Column 3: Integration range of the 9.7 $\mu$m silicate feature. Column 4: Total integrated flux of the 9.7 $\mu$m silicate feature. Column 5: total equivalent width for the 9.7 $\mu$m silicate feature. Column 6: $S_{9.7\,\mu\mathrm{m}}\,$; see \autoref{sec:s97Analysis}.}
\end{deluxetable*}
\capstarttrue

%% file: tab5.tex
\capstartfalse
\begin{deluxetable*}{rccccc}
\tabletypesize{\scriptsize}
\tablecolumns{6}
\tablewidth{0pt}
\tablecaption{$S_{9.7\,\mu\mathrm{m}}\,$ Continuum Parameters and Measured Values of the ULIRG/PG QSO Sample
\label{tab:ulirgSilicateProperties}}
\tablehead{
\colhead{(Name)}    & \colhead{Pivot Points} & \colhead{Integration Range} & \colhead{Flux$_{S_{9.7\,\mu\mathrm{m}}}$} & \colhead{EQW$_{S_{9.7\,\mu\mathrm{m}}}$} &  \colhead{$S_{9.7\,\mu\mathrm{m}}\,$} \\
\colhead{}    & \colhead{($\mu$m)} &  \colhead{($\mu$m)} & \colhead{(Jy $\mu$m)} & \colhead{($\mu$m)}   & \colhead{}\\
\cline{1-6}\\
\colhead{(1)}    & \colhead{(2)} & \colhead{(3)}   & \colhead{(4)}    & \colhead{(5)} & \colhead{(6)}
}
\startdata
07251$-$0248	&	5.2, 14.0, 27.0, 31.5	&	8.45, 12.7	&	0.21	&	3.4	&	2.37\\ 
09022$-$3615	&	5.2, 5.6, 14.0, 27.0, 31.5	&	8.3, 12.6	&	0.43	&	1.84	&	1.22\\ 
13120$-$5453	&	5.2, 5.6, 14.0, 27.0, 31.5	&	8.7, 12.0	&	0.5	&	1.72	&	1.49\\ 
19542+1110	&	5.2, 5.6, 14.0, 27.0, 31.5	&	8.74,11.14	&	0.05	&	1.12	&	0.86\\
F00509+1225	&	5.2, 5.6, 14.0, 27.0, 31.5	&	7.8, 14.0	&	0.31	&	0.82	&	$-$0.24\\ 
F01572+0009	&	5.2, 5.6, 14.0, 27.0, 31.5	&	8.1, 14.0	&	0.06	&	0.75	&	$-$0.17\\ 
F05024$-$1941	&	5.2, 5.6, 10.8, 18.5	&	7.8, 12.6	&	0.01	&	1.65	&	0.22\\
F05189$-$2524	&	5.2, 5.6, 14.0, 27.0, 31.5	&	8.37, 14.0	&	0.54	&	1.01	&	0.37\\ 
F07598+6508	&	5.2, 5.6, 14.0, 27.0, 31.5	&	7.2, 14.0	&	0.11	&	0.49	&	$-$0.15\\ 
F08572+3915	&	5.2, 5.6, 7.8, 14.0, 27.0, 31.5	&	7.8, 13.0	&	2.55	&	3.39	&	4.00\\ 
F09320+6134	&	5.2, 5.6, 14.0, 27.0, 31.5	&	8.65, 12.1	&	0.33	&	1.91	&	1.58\\ 
F10565+2448	&	5.2, 5.6, 14.0, 27.0, 31.5	&	8.76, 11.17	&	0.16	&	1.21	&	1.01\\ 
F11119+3257	&	5.2, 5.6, 7.8, 14.0, 27.0, 31.5	&	7.8, 12.6	&	0.17	&	1.21	&	0.62\\ 
F12072$-$0444	&	5.2, 5.6, 14.0, 27.0, 31.5	&	8.1, 12.6	&	0.22	&	2.36	&	1.41\\ 
F12112+0305	&	5.2, 5.6, 14.0, 27.0, 31.5	&	8.6, 12.4	&	0.13	&	2.42	&	1.58\\ 
F12243$-$0036	&	5.2, 5.6, 7.8, 14.0, 27.0, 31.5	&	8.15, 12.8	&	4.73	&	3.95	&	4.03\\ 
F12265+0219	&	5.2, 5.6, 14.0, 27.0, 31.5	&	8.8, 14.0	&	0.08	&	0.26	&	$-$0.06\\ 
F12540+5708	&	5.2, 5.6, 14.0, 27.0, 31.5	&	8.4, 12.6	&	2.62	&	1.49	&	0.62\\ 
F13305$-$1739	&	\nodata	&	\nodata	&	\nodata	&	\nodata	&	\nodata\\
F13428+5608	&	5.2, 5.6, 14.0, 27.0, 31.5	&	8.2, 12.6	&	0.73	&	2.77	&	2.02\\ 
F13451+1232	&	5.2, 5.6, 7.8, 14.0, 27.0, 31.5	&	7.8, 12.2	&	0.06	&	0.61	&	0.35\\ 
F14348$-$1447	&	5.2, 5.6, 14.0, 27.0, 31.5	&	8.3, 12.4	&	0.15	&	2.66	&	1.93\\ 
F14378$-$3651	&	5.2, 5.6, 14.0, 27.0, 31.5	&	8.7, 12.1	&	0.06	&	1.76	&	1.55\\ 
F14394+5332	&	5.2, 5.6, 14.0, 27.0, 31.5	&	8.5, 12.3	&	0.18	&	2.26	&	1.71\\ 
F15206+3342	&	5.2, 5.6, 14.0, 27.0, 31.5	&	9.0, 11.1	&	0.02	&	0.47	&	0.29\\ 
F15250+3608	&	5.2, 5.6, 7.8, 14.0, 27.0, 31.5	&	7.8, 14.0	&	0.94	&	3.54	&	3.42\\ 
F15327+2340	&	5.2, 5.6, 7.8, 14.0, 27.0, 31.5	&	7.8, 12.6	&	2.87	&	3.57	&	3.49\\ 
F15462$-$0450	&	5.2, 5.6, 14.0, 27.0, 31.5	&	8.6, 12.4	&	0.06	&	0.78	&	0.29\\ 
F16504+0228	&	5.2, 5.6, 14.0, 27.0, 31.5	&	8.65, 12.2	&	0.72	&	2.06	&	1.54\\ 
F17208$-$0014	&	5.2, 5.6, 14.0, 27.0, 31.5	&	8.6, 12.4	&	0.37	&	2.38	&	1.96\\ 
F19297$-$0406	&	5.2, 5.6, 14.0, 27.0, 31.5	&	8.6, 12.5	&	0.12	&	2.48	&	1.60\\ 
F20551$-$4250	&	5.2, 5.6, 7.8, 14.0, 27.0, 31.5	&	7.8, 14.0	&	1.52	&	3.43	&	3.09\\ 
F22491$-$1808	&	5.2, 5.6, 14.0, 27.0, 31.5	&	8.65, 12.4	&	0.07	&	2.2	&	1.14\\ 
F23128$-$5919	&	5.2, 5.6, 14.0, 27.0, 31.5	&	8.65, 12.2	&	0.23	&	1.45	&	0.88\\ 
F23233+2817	&	\nodata	&	\nodata	&	\nodata	&	\nodata	&	\nodata\\
F23365+3604	&	5.2, 5.6, 14.0, 27.0, 31.5	&	8.65, 12.4	&	0.17	&	2.49	&	1.78\\ 
F23389+0300	&	\nodata	&	\nodata	&	\nodata	&	\nodata	&	\nodata\\
PG 1126$-$041	&	\nodata	&	\nodata	&	\nodata	&	\nodata	&	\nodata\\
PG 1351+640	&	5.2, 5.6, 14.0, 27.0, 31.5	&	8.1, 14.0	&	0.24	&	2.31	&	$-$0.58\\
PG 1440+356	&	\nodata	&	\nodata	&	\nodata	&	\nodata	&	\nodata\\
PG 1613+658	&	5.2, 5.6, 14.0, 27.0, 31.5	&	7.2, 14.0	&	0.03	&	0.38	&	$-$0.01\\
PG 2130+099	&	5.2, 5.6, 14.0, 27.0, 31.5	&	7.75, 14.0	&	0.05	&	0.29	&	0.10\\
Z11598$-$0114	&	\nodata	&	\nodata	&	\nodata	&	\nodata	&	\nodata
\enddata
\tablecomments{Column 1: galaxy name. Column 2: Pivot points in microns used for the continuum interpolation. Column 3: Integration range of the 9.7 $\mu$m silicate feature. Column 4: Total integrated flux of the 9.7 $\mu$m silicate feature. Column 5: total equivalent width for the 9.7 $\mu$m silicate feature. Column 6: $S_{9.7\,\mu\mathrm{m}}\,$; see \autoref{sec:s97Analysis}.}
\end{deluxetable*}
\capstarttrue

%% file: tab6.tex
\capstartfalse
\begin{deluxetable*}{l c c c c c c c}
\tablecolumns{8}
\tabletypesize{\scriptsize}
\tablewidth{0pt}
\tablecaption{Results from Statistical Analyses of Host Galaxy Properties
\label{tab:statResultsHostGalaxy}}
\tablehead{
\colhead{Parameter}    & \colhead{Number of Objects}    &  \colhead{$\rho_s$} & \colhead{P$_\rho$}   & \colhead{$\tau$} & \colhead{P$_\tau$} & \colhead{$r$} & \colhead{P$_r$} \\
\cline{1-8}\\
\colhead{(1)}    & \colhead{(2)} & \colhead{(3)}   & \colhead{(4)}    & \colhead{(5)} & \colhead{(6)} & \colhead{(7)} & \colhead{(8)}
}
\startdata
log $M_*$ - EW$_\mathrm{OH}$ (ULIRGs Only)	&	21	&	$-$0.49	&	2.5e$-$02	&	$-$0.34	&	3.3e$-$02	&	$-$0.56	&	\bf{8.4e$-$03}\\ 
log $M_*$ - EW$_\mathrm{OH}$ (BAT AGN only)	&	18	&	0.04	&	8.8e$-$01	&	0.04	&	8.2e$-$01	&	$-$0.00	&	9.9e$-$01\\ 
log $M_*$ - EW$_\mathrm{OH}$ (Combined Sample)	&	39	&	0.13	&	4.3e$-$01	&	0.09	&	4.0e$-$01	&	0.13	&	4.2e$-$01\\ 
& & & & & \\
$\alpha_{\mathrm{AGN}}$ - EW$_\mathrm{OH}$ (ULIRGs Only)	&	37	&	$-$0.42	&	1.1e$-$02	&	$-$0.29	&	1.0e$-$02	&	$-$0.47	&	\bf{3.7e$-$03}\\ 
$\alpha_{\mathrm{AGN}}$ - EW$_\mathrm{OH}$ (BAT AGN only)	&	38	&	$-$0.11	&	5.2e$-$01	&	$-$0.08	&	5.0e$-$01	&	$-$0.27	&	1.0e$-$01\\ 
$\alpha_{\mathrm{AGN}}$ - EW$_\mathrm{OH}$ (Combined Sample)	&	75	&	$-$0.42	&	\bf{1.8e$-$04}	&	$-$0.29	&	\bf{2.3e$-$04}	&	$-$0.44	&	\bf{7.0e$-$05}\\ 
& & & & & \\
log $L_{\mathrm{AGN}}$ - EW$_\mathrm{OH}$ (ULIRGs Only)	&	37	&	$-$0.25	&	1.3e$-$01	&	$-$0.19	&	9.7e$-$02	&	$-$0.31	&	6.4e$-$02\\ 
log $L_{\mathrm{AGN}}$ - EW$_\mathrm{OH}$ (BAT AGN only)	&	42	&	0.03	&	8.5e$-$01	&	0.02	&	8.6e$-$01	&	$-$0.03	&	8.7e$-$01\\ 
log $L_{\mathrm{AGN}}$ - EW$_\mathrm{OH}$ (Combined Sample)	&	79	&	0.26	&	2.0e$-$02	&	0.18	&	2.1e$-$02	&	0.25	&	2.4e$-$02\\ 
\enddata
\tablecomments{Column 1: quantities considered for the statistical test. Column 2: number of objects in which OH 119 $\mu$m is detected. Column 3: Spearman rank order correlation coefficient. Column 4: null probability of the Spearman rank order correlation coefficient. Column 5: Kendall's correlation coefficient. Column 6: null probability of Kendall's correlation. Column 7: Pearson's linear correlation coefficient. Column 8: Two-tail area probability of Pearson's linear correlation. Null probabilities $\lesssim 10^{-3}$ (shown in bold-faced characters) indicate statistically significant trends.}
\end{deluxetable*}
\capstarttrue

%% file: tab7.tex
\capstartfalse 
\begin{deluxetable*}{l c c c c c c c}
\tablecolumns{8}
\tabletypesize{\scriptsize}
\tablewidth{0pt}
\tablecaption{Results from Statistical Analyses of the Kinematics
\label{tab:statResults}}
\tablehead{
\colhead{Parameter}    & \colhead{Number of Objects}    &  \colhead{$\rho_s$} & \colhead{P$_\rho$}   & \colhead{$\tau$} & \colhead{P$_\tau$} & \colhead{$r$} & \colhead{P$_r$} \\
\cline{1-8}\\
\colhead{(1)}    & \colhead{(2)} & \colhead{(3)}   & \colhead{(4)}    & \colhead{(5)} & \colhead{(6)} & \colhead{(7)} & \colhead{(8)}
}
\startdata
\cutinhead{ULIRGs Only}
$\log M_*-v_{50}$ 	&	18	&	-0.15	&	5.6e-01	&	-0.15	&	3.8e-01	&	-0.04	&	8.7e-01\\ 
$\log M_*-v_{84}$ 	&	18	&	-0.10	&	7.0e-01	&	-0.08	&	6.4e-01	&	0.01	&	9.6e-01\\ 
&	&	&	&	&	&	&	\\ 
$\log SFR-v_{50}$ 	&	32	&	0.28	&	1.2e-01	&	0.22	&	8.2e-02	&	0.06	&	7.6e-01\\ 
$\log SFR-v_{84}$ 	&	32	&	0.19	&	2.9e-01	&	0.14	&	2.4e-01	&	0.07	&	7.1e-01\\ 
&	&	&	&	&	&	&	\\ 
sSFR $-\,v_{50}$ 	&	18	&	0.37	&	1.3e-01	&	0.26	&	1.3e-01	&	0.54	&	2.2e-02\\ 
sSFR $-\,v_{84}$ 	&	18	&	0.26	&	3.0e-01	&	0.21	&	2.3e-01	&	0.57	&	1.4e-02\\ 
&	&	&	&	&	&	&	\\ 
$\alpha_{\mathrm{AGN}}-v_{50}$ 	&	32	&	-0.52	&	\bf{2.4e-03}	&	-0.37	&	\bf{3.2e-03}	&	-0.48	&	\bf{5.5e-03}\\ 
$\alpha_{\mathrm{AGN}}-v_{84}$ 	&	32	&	-0.46	&	\bf{7.4e-03}	&	-0.31	&	1.2e-02	&	-0.46	&	\bf{7.6e-03}\\ 
&	&	&	&	&	&	&	\\ 
$\log L_{\mathrm{AGN}}-v_{50}$ 	&	32	&	-0.50	&	\bf{3.3e-03}	&	-0.37	&	\bf{2.8e-03}	&	-0.56	&	\bf{8.0e-04}\\ 
$\log L_{\mathrm{AGN}}-v_{84}$ 	&	32	&	-0.44	&	1.1e-02	&	-0.32	&	\bf{9.8e-03}	&	-0.54	&	\bf{1.6e-03}\\
\cutinhead{BAT AGN Only}
$\log M_*-v_{50}$ 	&	6	&	-0.23	&	6.6e-01	&	-0.14	&	7.0e-01	&	-0.34	&	5.1e-01\\ 
$\log M_*-v_{84}$ 	&	6	&	0.09	&	8.7e-01	&	-0.07	&	8.5e-01	&	0.27	&	6.1e-01\\ 
&	&	&	&	&	&	&	\\ 
$\log SFR-v_{50}$	&	13	&	-0.11	&	7.3e-01	&	-0.11	&	6.2e-01	&	-0.04	&	8.9e-01\\ 
$\log SFR-v_{84}$ 	&	13	&	0.24	&	4.4e-01	&	0.17	&	4.2e-01	&	-0.02	&	9.5e-01\\ 
&	&	&	&	&	&	&	\\ 
sSFR $-\,v_{50}$	&	6	&	0.32	&	5.4e-01	&	0.28	&	4.4e-01	&	0.17	&	7.4e-01\\ 
sSFR $-\,v_{84}$ 	&	6	&	0.54	&	2.7e-01	&	0.33	&	3.5e-01	&	0.34	&	5.1e-01\\ 
&	&	&	&	&	&	&	\\ 
$\alpha_{\mathrm{AGN}}-v_{50}$ &	13	&	-0.50	&	8.4e-02	&	-0.32	&	1.3e-01	&	-0.50	&	8.1e-02\\ 
$\alpha_{\mathrm{AGN}}-v_{84}$ &	13	&	-0.69	&	\bf{9.4e-03}	&	-0.50	&	1.7e-02	&	-0.44	&	1.3e-01\\ 
&	&	&	&	&	&	&	\\ 
$\log L_{\mathrm{AGN}}-v_{50}$ 	&	14	&	-0.27	&	3.6e-01	&	-0.18	&	3.7e-01	&	-0.39	&	1.7e-01\\ 
$\log L_{\mathrm{AGN}}-v_{84}$	&	14	&	-0.36	&	2.1e-01	&	-0.27	&	1.9e-01	&	-0.29	&	3.2e-01\\
\cutinhead{Combined Sample}
$\log M_*-v_{50}$ 	&	24	&	-0.58	&	\bf{2.8e-03}	&	-0.44	&	\bf{2.5e-03}	&	-0.13	&	5.5e-01\\ 
$\log M_*-v_{84}$ 	&	24	&	-0.53	&	\bf{7.9e-03}	&	-0.38	&	\bf{9.9e-03}	&	-0.08	&	7.0e-01\\ 
&	&	&	&	&	&	&	\\ 
$\log SFR-v_{50}$ 	&	45	&	-0.34	&	2.3e-02	&	-0.21	&	3.9e-02	&	-0.44	&	\bf{2.3e-03}\\ 
$\log SFR-v_{84}$ 	&	45	&	-0.30	&	4.3e-02	&	-0.18	&	7.8e-02	&	-0.40	&	\bf{6.8e-03}\\ 
&	&	&	&	&	&	&	\\ 
sSFR $-\,v_{50}$ 	&	24	&	0.13	&	5.6e-01	&	0.10	&	4.8e-01	&	0.34	&	1.1e-01\\ 
sSFR $-\,v_{84}$ 	&	24	&	0.10	&	6.3e-01	&	0.09	&	5.2e-01	&	0.38	&	6.7e-02\\ 
&	&	&	&	&	&	&	\\ 
$\alpha_{\mathrm{AGN}}-v_{50}$ &	45	&	-0.04	&	8.1e-01	&	-0.03	&	7.8e-01	&	-0.08	&	6.1e-01\\ 
$\alpha_{\mathrm{AGN}}-v_{84}$ &	45	&	-0.09	&	5.6e-01	&	-0.06	&	5.7e-01	&	-0.12	&	4.5e-01\\ 
&	&	&	&	&	&	&	\\ 
$\log L_{\mathrm{AGN}}-v_{50}$ 	&	46	&	-0.70	&	\bf{6.8e-08}	&	-0.51	&	\bf{5.7e-07}	&	-0.67	&	\bf{3.3e-07}\\ 
$\log L_{\mathrm{AGN}}-v_{84}$ 	&	46	&	-0.61	&	\bf{8.2e-06}	&	-0.45	&	\bf{8.9e-06}	&	-0.63	&	\bf{2.9e-06}
\enddata
\tablecomments{Column 1: quantities considered for the statistical test. Column 2: number of objects in which OH 119 $\mu$m is detected in either redshifted or blueshifted absorption. Column 3: Spearman rank order correlation coefficient. Column 4: null probability of the Spearman rank order correlation coefficient. Column 5: Kendall's correlation coefficient. Column 6: null probability of Kendall's correlation. Column 7: Pearson's linear correlation coefficient. Column 8: Two-tail area probability of Pearson's linear correlation. Null probabilities $\lesssim 10^{-3}$ (shown in bold-faced characters) indicate statistically significant trends.}
\end{deluxetable*}
\capstarttrue